\documentclass[11pt]{article}

\usepackage{amsmath,scalefnt}
\usepackage{amssymb}
\usepackage{amstext}
\usepackage{latexsym}
\usepackage{color}
\usepackage{graphicx}
\usepackage{epsf}
\usepackage{epsfig}
\usepackage{subfigure}
\usepackage{rotating}
\usepackage{curves}
\usepackage{epic}
\usepackage{amsthm}

\usepackage{amsfonts}
\usepackage[T1]{fontenc}
\usepackage[latin1]{inputenc}
\usepackage{authblk}
\usepackage{tikz}
\usetikzlibrary{shapes,arrows}
\usepackage{graphics}
\usepackage{verbatim}
\usepackage{color}
\usepackage{hyperref}
\usepackage{mathtools}
\usepackage{epstopdf}
\allowdisplaybreaks[1]

\newcommand{\be}{\begin{equation}}
\newcommand{\en}{\end{equation}}

\newcommand{\1}{1 \!\! 1}

\newtheorem{thm}{Theorem}
\newtheorem{cor}[thm]{Corollary}

\newtheorem{defi}{Definition}[section]
\newtheorem{lem}[defi]{Lemma}

\newtheorem{Theo}{Theorem}[section]

\newcommand{\bedefin}{\begin{defi}}
\newcommand{\findefi}{\end{defi} \medskip}

\newcommand{\betheo}{\begin{theorem}$\!\!${\bf \,\,\,}}
\newcommand{\entheo}{\end{theorem}}
\newcommand{\enth}{\end{theorem}}

\newcommand{\becor}{\begin{cor}$\!\!${\bf .}}
\newcommand{\encor}{\end{cor}}

\newcommand{\belem}{\begin{lem}$\!\!${\bf }}
\newcommand{\enlem}{\end{lem}}

\newcommand{\bea}{\begin{eqnarray}}
\newcommand{\ena}{\end{eqnarray}}

\newcommand{\beano}{\begin{eqnarray*}}
\newcommand{\enano}{\end{eqnarray*}}

\newcommand{\bee}{\begin{enumerate}}
\newcommand{\ene}{\end{enumerate}}

\newcommand{\bei}{\begin{itemize}}
\newcommand{\eni}{\end{itemize}}

\newcommand{\betab}{\begin{tabular}}
\newcommand{\entab}{\end{tabular}}

\newcommand{\bd}{\begin{displaymath}}

\newcommand{\p}{\mbox{${\mathcal P}^{\uparrow}_{+}(1,1)$}}

\catcode `\@=11 \@addtoreset{equation}{section}

\catcode `\@=12

\begin{document}

\title{Coadjoint Orbits and Wigner functions of (1+1)-Extended Affine Galilei Group and Galilei Group}
\author{S. Hasibul Hassan Chowdhury$^\dag$ \\
  and S. Twareque Ali$^{\dag\dag}$\\
\medskip
\small{Department of Mathematics and Statistics,\\ Concordia University, Montr\'eal, Qu\'ebec, Canada H3G 1M8}\\

{\footnotesize $^\dag$e-mail: schowdhury@mathstat.concordia.ca\\
 $^{\dag\dag}$e-mail: stali@mathstat.concordia.ca}}

\date{\today}

\maketitle

\begin{abstract}
We study the coadjoint orbits of the noncentrally extended (1+1)-affine Galilei group  and compute the relevant Wigner functions built on them explicitly. We consider the centrally extended (1+1)-Galilei group and study its coadjoint orbits in the second half of the paper. We also compute the Wigner functions built on the corresponding coadjoint orbits subsequently. Finally, a comparative study of the structure of the coadjoint orbits and corresponding Wigner functions between the extended (1+1)-affine Galilei group  and the centrally extended (1+1)-Galilei group is presented along with possible physical interpretations.\\
\rule{1.5in}{0.02in}
\end{abstract}

%\tableofcontents

\section{Introduction}\label{sec:intro}
(1+1)- Galilei group $\mathcal{G}_{0}$ is the Kinematical group of non relativistic spacetime of dimension (1+1). In \cite{mahara} an extension of (n+1)-Galilei group by the two dimensional dilation group $\mathcal{D}_{2}$ (independent space and time dilations) has been considered for $n\geqslant 3$. The resulting extended group is referred to as the (n+1)-affine Galilei group in the literature. We follow the similar construction to obtain (1+1)-affine Galilei group $\mathcal{G}_{\hbox{\tiny{aff}}}$.

This group has profound significance in signal analysis and image processing \cite{affgal}. But this group does not seem to have any quantum mechanical features associated with it. In order to have a well defined quantum mechanical feature we have to consider the projective representation of the underlying group. In other words, we have to find a nontrivial central extension of the given group and consider the true reprsentations of the centrally (nontrivial) extended group. But as in the higher dimensional case (3 or more) \cite{mahara}, one could show that the straightforward central extension of $\mathcal{G}_{\hbox{\tiny{aff}}}$ fails to generate the mass of the nonrelativistic spinless particle under the stated symmetry. This problematic feature was remedied by the two step construction of a noncentral extension of (1+1)-affine Galilei group $\mathcal{G}_{\hbox{\tiny{aff}}}$. First taking the central extension of the (1+1)-Galilei group $\mathcal{G}_{0}$, and then taking the semidirect product of the resulting extended group $\mathcal{G}^{m}$ with the two dimensional dilation group $\mathcal{D}_{2}$. In this way, we arrive at the group $\mathcal{G}^{m}_{\hbox{\tiny{aff}}}=\mathcal{G}^{m}\rtimes\mathcal{D}_{2}$. It is to be noted that the group so obtained is a noncentral extension of the (1+1)-affine Galilei group $\mathcal{G}_{\hbox{\tiny{aff}}}$.

\section{Wigner functions of (1+1)-extended affine Galilei group}

  The group $\mathcal{G}^{m}_{\hbox{\tiny{aff}}}$ is defined by the following continuous transformation
\begin{eqnarray*}
x&\mapsto& e^{\sigma}x+e^{\tau}vt+a\\
t&\mapsto& e^{\tau}t+b
\end{eqnarray*}
where in addition to the parameters $(b,a,v)$ of the (1+1) dimensional Galilei group $\mathcal{G}_{0}$ we have two more parameters $\sigma,\tau\in\mathbb{R}$ representing independent space and time dilations respectively. So a generic element $g$ of the (1+1) dimensional affine Galilei group is represented as $(b,a,v,\sigma,\tau)$.

On the other hand, a generic element of (1+1) dimensional extended Galilei group $\mathcal{G}^{m}_{\hbox{\tiny{aff}}}$ is represented as $(\theta,b,a,v,\sigma,\tau)$ obeying the following group composition law
\begin{eqnarray*}
\lefteqn{(\theta, b, a, v,\sigma,\tau)(\theta^\prime,b^\prime,a^\prime,v^\prime,\sigma^\prime,\tau^\prime)}\\
&&=\theta+e^{2\sigma-\tau}\theta^{\prime}+m[e^{\sigma}va^{\prime}+\frac{1}{2}e^{\tau}v^{2}b^{\prime}],
b+e^{\tau}b^{\prime},a+e^{\tau}b^{\prime}v+e^{\sigma}a^{\prime}, v+e^{\sigma-\tau}v^{\prime},\\
&&\hspace{4in}\sigma+\sigma^{\prime},\tau+\tau^{\prime})
\end{eqnarray*}
In this section we will study various coadjoint orbits of $\mathcal{G}^{m}_{\hbox{\tiny{aff}}}$ and develop the required tools to compute the Wigner functions built on them.

\subsection{Dual orbits of (1+1)-extended affine Galilei group}\label{subsec:dual-orbt-aff-gal}
An element $(\theta,b,a,v,\sigma,\tau)$ of $\mathcal{G}^{m}_{\hbox{aff}}$ can be represented by the following matrix
\begin{equation}\label{eq:matrep-ex-aff-gal}
(\theta,b,a,v,\sigma,\tau)=\begin{bmatrix}e^{\sigma}&ve^{\tau}&0&a\\0&e^{\tau}&0&b\\mve^{\sigma}&\frac{1}{2}mv^{2}e^{\tau}&e^{2\sigma-\tau}&\theta\\0&0&0&1\end{bmatrix}
\end{equation}
where the group multiplication now reduces to matrix multiplication for the matrices given by (\ref{eq:matrep-ex-aff-gal}). It can easily be seen that $\mathcal{T}={(\theta,b,a,0,0,0)}\sim\mathbb{R}^{3}$ and $\mathcal{V}={(0,0,0,v,0,0)}\sim\mathbb{R}$ are abelian subgroups of $\mathcal{G}^{m}_{\hbox{\tiny{aff}}}$. In terms of these two abelian subgroups, $\mathcal{G}^{m}_{\hbox{\tiny{aff}}}$ can be written as $\mathcal{G}^{m}_{\hbox{\tiny{aff}}}=\mathcal{T}\rtimes(\mathcal{V}\rtimes\mathbb{R}^{2})$.
Now we proceed to find the dual orbits of $\mathcal{G}^{m}_{\hbox{\tiny{aff}}}$ under $H=\mathcal{V}\rtimes\mathbb{R}^{2}$ in $\mathcal{T}^{*}\sim\hat{\mathbb{R}}^3$. If we denote by $(q,E,p)$  a generic element of $\mathcal{T}^{*}$ then the action of $(v,\sigma,\tau)\in H$ on $(q,E,p)$ is found to be
\begin{equation}\label{eq:dualact-ex-aff-gal}
(v,\sigma,\tau)(q,E,p)=(e^{\tau-2\sigma}q,e^{-\tau}E+e^{-\sigma}pv+\frac{1}{2}qme^{\tau-2\sigma}v^{2},e^{-\sigma}p+e^{\tau-2\sigma}qmv).
\end{equation}

The set of all possible triples $(q^{\prime},E^{\prime},p^{\prime})\in\mathbb{R}^3$ such that $(v,\sigma,\tau)(q,E,p)=(q^{\prime},E^{\prime},p^{\prime})$, form the dual orbit due to the element $(q,E,p)$ under $H$ in $\mathbb{R}^3$. So we have to solve the following system of equations for $(q^{\prime},E^{\prime},p^{\prime})$
\begin{eqnarray}\label{eq:dualacteqn-ex-aff-gal}
q^{\prime}&=&e^{\tau-2\sigma}q\nonumber\\
E^{\prime}&=& e^{-\tau}E+e^{-\sigma}pv+\frac{1}{2}qme^{\tau-2\sigma}v^{2}\\
p^{\prime}&=& e^{-\sigma}p+e^{\tau-2\sigma}qmv\nonumber
\end{eqnarray}
From (\ref{eq:dualacteqn-ex-aff-gal}), for nonzero values of $q$, it follows immediately that
\begin{eqnarray*}
e^{2\sigma}\frac{q^{\prime}}{q}&=&e^{\tau}\\
E^{\prime}-\frac{p^{\prime 2}}{2q^{\prime}m}&=&e^{-\tau}(E-\frac{p^2}{2mq})
\end{eqnarray*}
which in turn reflects the fact that the signs of both $q$ and $E-\frac{p^2}{2qm}$ are invariants on the same orbit. For different values of $q$, $p$, $E$, and $E-\frac{p^2}{2qm}$ we have eleven possible orbits as outlined in the following table.

\begin{table} [h]
\caption{All possible orbits of $\mathcal{G}^{m}_{\hbox{\tiny{aff}}}$ in $\mathbb{R}^3$ under $H=\mathcal{V}\rtimes\mathbb{R}^2$ }
\label{tab:tablefirst}
\centering
\begin{tabular}{c|c|c|c|c}
\hline
Orbits & $q$ & $p$ & $E$ & $E-\frac{p^2}{2qm}$\\
\hline
\hline
$\hat{\mathcal{O}}_{1}$ & $>0$ & $-$ & $-$ & $>0$ \\ % inserting body of the table
$\hat{\mathcal{O}}_{2}$ & $>0$ & $-$ & $-$ & $<0$ \\
$\hat{\mathcal{O}}_{3}$ & $<0$ & $-$ & $-$ & $<0$ \\
$\hat{\mathcal{O}}_{4}$ & $<0$ & $-$ & $-$ & $>0$ \\
$\hat{\mathcal{O}}_{5}$ & $>0$ & $-$ & $-$ & $=0$ \\
$\hat{\mathcal{O}}_{6}$ & $<0$ & $-$ & $-$ & $=0$ \\
$\hat{\mathcal{O}}_{7}$ & $=0$ & $>0$ & $\in\mathbb{R}$ & $-$ \\
$\hat{\mathcal{O}}_{8}$ & $=0$ & $<0$ & $\in\mathbb{R}$ & $-$ \\
$\hat{\mathcal{O}}_{9}$ & $=0$ & $=0$ & $>0$ & $-$ \\
$\hat{\mathcal{O}}_{10}$ & $=0$ & $=0$ & $<0$ & $-$ \\
$\hat{\mathcal{O}}_{11}$ & $=0$ & $=0$ & $=0$ & $-$ \\
\hline
\end{tabular}
\end{table}

\begin{figure}[thb]
\centering
\includegraphics[width=6cm]{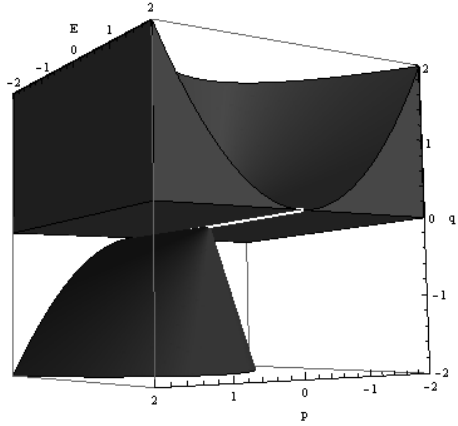}
\caption{The four open free orbits of (1+1) dimensional extended affine Galilei group: the hollow region in the upper half-region ($q>0$) represents $\hat{\mathcal{O}}_1$, and the filled region in the same half-region  represents $\hat{\mathcal{O}}_{2}$. Similarly the filled region in the lower half-region ($q<0$) represents $\hat{\mathcal{O}}_3$ and the corresponding hollow region there represents $\hat{\mathcal{O}}_4$.}
\label{fig:figfirst}
\end{figure}

\begin{figure}[thb]
\centering
\includegraphics[width=8cm]{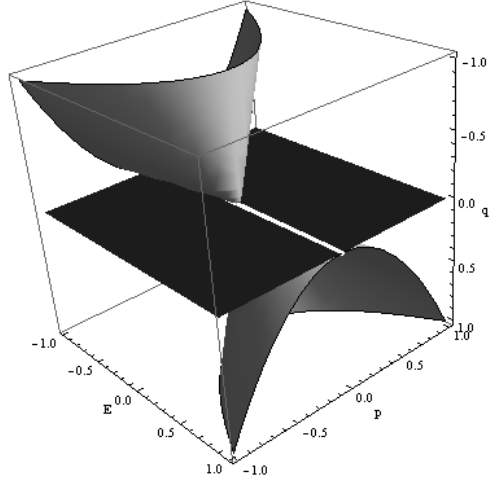}
\caption{The degenerate orbits of (1+1) dimensional affine Galilei group: the two dimensional surface in the upper half-region ($q>0$) represents $\hat{\mathcal{O}}_5$ and the one underneath ($q<0$) represents $\hat{\mathcal{O}}_6$. Also, the plane $q=0$ is the disjoint union of the other degenerate orbits $\hat{\mathcal{O}}_{7},\hat{\mathcal{O}}_{8}, \hat{\mathcal{O}}_{9},\hat{\mathcal{O}}_{10}$, and $\hat{\mathcal{O}}_{11}$.}
\label{fig:figsecond}
\end{figure}

The first four orbits $\hat{\mathcal{O}}_{1},\hat{\mathcal{O}}_{2}, \hat{\mathcal{O}}_{3}$, and $\hat{\mathcal{O}}_{4}$ of Table 1 listed above are three dimensional regions depicted in Figure \ref{fig:figfirst}. $\hat{\mathcal{O}}_{5}$ and $\hat{\mathcal{O}}_{6}$ are the two dimensional surfaces described in Figure \ref{fig:figsecond}, while $\hat{\mathcal{O}}_{7}$ and $\hat{\mathcal{O}}_{8}$ are the two half-planes $\mathbb{R}\times\mathbb{R}^{>0}$ ($p>0$) and $\mathbb{R}\times\mathbb{R}^{<0}$ ($p<0$) respectively due to $q=0$. Now, $\hat{\mathcal{O}}_{9}$ and $\hat{\mathcal{O}}_{10}$ represent the positive $E$-axis ($\mathbb{R}^{>0}$) and the negative $E$-axis ($\mathbb{R}^{<0}$) respectively. Finally, $\hat{\mathcal{O}}_{11}$ stands for the origin ($q=p=E=0$). It is interesting to see that the 3 dimensional non-degenerate orbits are disjoint and separated from one another lying in the same half regions (determined by $q>0$ or $q<0$) by the degenerate orbits $\hat{\mathcal{O}}_{5}$ and $\hat{\mathcal{O}}_{6}$ of Table \ref{tab:tablefirst}. Also, two of them lying in opposite half regions are separated by the two dimensional plane corresponding to $q=0$ ($\cup_{i=7}^{11}\hat{\mathcal{O}}_{i}$). Now, it is obvious that the first four orbits are open sets in $\mathbb{R}^3$. And it is easily verified using equation (3) that the set of all $(v,\sigma,\tau)\in\mathbb{R}^3$ such that $(v,\sigma,\tau)(q,E,p)=(q,E,p)$ is trivial, i.e, the element (0,0,0), which in turn implies that the stabilizer subgroup is trivial. So, the first four orbits are indeed open free and the two-dimensional surfaces in Figure 2 and the $q=0$ plane separate these open free orbits.

\subsection{Haar measures for the (1+1)-extended affine Galilei group and the corresponding modular function}\label{subsec:haar-msr-ex-aff-gal}

Consider a fixed group element $g_0$ and let it act on another element $g$ from the left to obtain the following
\begin{equation*}
g_{0}g=\begin{bmatrix}e^{\sigma^{\prime}}&v^{\prime}e^{\tau^{\prime}}&0&a^{\prime}\\0&e^{\tau^{\prime}}&0&b^{\prime}\\mv^{\prime}e^{\sigma^{\prime}}&\frac{1}{2}mv^{\prime 2}e^{\tau^{\prime}}&e^{2\sigma^{\prime}-\tau^{\prime}}&\theta^{\prime}\\0&0&0&1\end{bmatrix},
\end{equation*}
where
\begin{eqnarray}\label{eq:left-haar-tranfm}
&&\sigma^{\prime}=\sigma+\sigma_{0},\; \tau^{\prime}=\tau+\tau_{0},\; v^{\prime}=ve^{\sigma_{0}-\tau_{0}}+v_{0},\;
a^{\prime}=e^{\sigma_{0}}a+v_{0}e^{\tau_{0}}b+a_{0},\\
&&b^{\prime}=e^{\tau_{0}}b+b_{0}\; \theta^{\prime}=mv_{0}e^{\sigma_{0}}a+\frac{1}{2}mv_{0}^{2}e^{\tau_{0}}b+e^{2\sigma_{0}-\tau_{0}}\theta+\theta_{0}\nonumber.
\end{eqnarray}
Therefore, under the left action of a fixed group element $g_{0}\equiv(\theta_0,b_0,a_0,v_0,\sigma_0,\tau_0)$, a generic group element $g\in\mathcal{G}^{m}_{\hbox{\tiny{aff}}}$ transforms as
\begin{equation*}
(\theta,b,a,v,\sigma,\tau)\mapsto(\theta^{\prime},b^{\prime},a^{\prime},v^{\prime},\sigma^{\prime},\tau^{\prime}).
\end{equation*}
obeying the system (\ref{eq:left-haar-tranfm}).
Now it follows that
\begin{eqnarray*}
&&d\sigma^{\prime}=d\sigma,\; d\tau^{\prime}=d\tau,\;dv^{\prime}=e^{\sigma_{0}-\tau_{0}}dv,\; da^{\prime}=e^{\sigma_{0}}da+v_{0}e^{\tau_{0}}db,\\
&&db^{\prime}=e^{\tau_{0}}db,\; d\theta^{\prime}=mv_{0}e^{\sigma_{0}}da+\frac{1}{2}mv_{0}^{2}e^{\tau_{0}}db+e^{2\sigma_{0}-\tau_{0}}d\theta.
\end{eqnarray*}
We can easily see from the above computation that
\begin{equation}\label{eq:eq-gvng-lft-haar-msr}
e^{-4\sigma^{\prime}+\tau^{\prime}}dv^{\prime}\wedge da^{\prime}\wedge db^{\prime}\wedge d\theta^{\prime}\wedge d\sigma^{\prime}\wedge d\tau^{\prime}=e^{-4\sigma+\tau}dv\wedge da\wedge db\wedge d\theta\wedge d\sigma \wedge d\tau
\end{equation}
Therefore, (\ref{eq:eq-gvng-lft-haar-msr}) suggests that $e^{-4\sigma+\tau}dv\wedge da\wedge db\wedge d\theta\wedge d\sigma\wedge d\tau$ is the left invariant Haar measure for the (1+1) dimensional extended affine Galilei group.

Now we act with $g_0$ on $g$ from the right:
\begin{equation*}
gg_{0}=\begin{bmatrix}e^{\sigma^{\prime}}&v^{\prime}e^{\tau^{\prime}}&0&a^{\prime}\\0&e^{\tau^{\prime}}&0&b^{\prime}\\mv^{\prime}e^{\sigma^{\prime}}&\frac{1}{2}mv^{\prime 2}e^{\tau^{\prime}}&e^{2\sigma^{\prime}-\tau^{\prime}}&\theta^{\prime}\\0&0&0&1\end{bmatrix},
\end{equation*}
with
\begin{eqnarray*}
&&\sigma^{\prime}=\sigma+\sigma_{0},\;\tau^{\prime}=\tau+\tau_{0},\;v^{\prime}=v+v_{0}e^{\sigma-\tau},\\
&&a^{\prime}=e^{\sigma}a_{0}+ve^{\tau}b_{0}+a,\;b^{\prime}=e^{\tau}b_{0}+b,\; \theta^{\prime}=mve^{\sigma}a_{0}+\frac{1}{2}mv^{2}e^{\tau}b_{0}+e^{2\sigma-\tau}\theta_{0}+\theta.
\end{eqnarray*}
So the Lebesgue measures along the group parameters transform in the following manner
\begin{eqnarray*}
&&d\sigma^{\prime}=d\sigma,\;d\tau^{\prime}=d\tau,\; dv^{\prime}=dv+v_{0}e^{\sigma-\tau}d\sigma-v_{0}e^{\sigma-\tau}d\tau,\\ &&da^{\prime}=a_{0}e^{\sigma}d\sigma+b_{0}e^{\tau}dv+b_{0}ve^{\tau}d\tau+da,\;db^{\prime}=b_{0}e^{\tau}d\tau+db,\\ &&d\theta^{\prime}=m(a_{0}e^{\sigma}+b_{0}ve^{\tau})dv+(ma_{0}ve^{\sigma}+2\theta_{0}e^{2\sigma-\tau})d\sigma
+(\frac{1}{2}mb_{0}v^{2}e^{\tau}-\theta_{0}e^{2\sigma-\tau})d\tau+d\theta.
\end{eqnarray*}
It follows immediately that
\begin{equation}\label{eq:eq-gvng-rght-haar-msr}
dv^{\prime}\wedge da^{\prime}\wedge db^{\prime}\wedge d\theta^{\prime}\wedge d\sigma^{\prime}\wedge d\tau^{\prime}=dv\wedge da\wedge db\wedge d\theta\wedge d\sigma \wedge d\tau
\end{equation}
Therefore, the right invariant Haar measure turns out to be the usual Lebesgue measure on the underlying group manifold, which is just $dv\wedge da\wedge db\wedge d\theta\wedge d\sigma \wedge d\tau$. Now, (\ref{eq:eq-gvng-lft-haar-msr}), together with (\ref{eq:eq-gvng-rght-haar-msr}), imply that the group $\mathcal{G}^{m}_{\hbox{\tiny{aff}}}$ is non-unimodular and  $e^{-4\sigma+\tau}$ is the required modular function for the underlying group.

\subsection{Lie algebraic aspects and the coadjoint action matrix of (1+1) dimensional extended affine Galilei group}\label{subsec:coadmtrx-ex-aff-gal}
We first observe  that our problem of the (1+1) dimensional extended affine Galilei group fits exactly into the framework of semidirect product groups discused in \cite{semidirect}. We are doing so because our ultimate goal is to construct Wigner map for $\mathcal{G}^{m}_{\hbox{\tiny{aff}}}$.
Following the matrix representation (\ref{eq:matrep-ex-aff-gal}) of a generic group element $g\equiv(\theta,b,a,v,\sigma,\tau)$ of $\mathcal{G}^{m}_{\hbox{\tiny{aff}}}$ we see that indeed $\mathcal{G}^{m}_{\hbox{\tiny{aff}}}=\mathbb{R}^{3}\rtimes H$, where $H=\left\{\begin{bmatrix}e^{\sigma}&ve^{\tau}&0\\0&e^{\tau}&0\\mve^{\sigma}&\frac{1}{2}mv^{2}e^{\tau}&e^{2\sigma-\tau}\end{bmatrix}|v,\sigma,\tau\in\mathbb{R}\right\}$ is a closed subgroup of $GL(3,\mathbb{R})$. And $(\theta,b,a)\in\mathbb{R}^3$.

From the above discussion we see that $g$ can be written as $(\vec{x},h)$ where $\vec{x}=\begin{pmatrix}a\\b\\\theta\end{pmatrix}\in\mathbb{R}^3$ and $h=\begin{bmatrix}e^{\sigma}&ve^{\tau}&0\\0&e^{\tau}&0\\mve^{\sigma}&\frac{1}{2}mv^{2}e^{\tau}&e^{2\sigma-\tau}\end{bmatrix}\in H$. And the semidirect product is given by
\begin{equation*}
(\vec{x}_{1},h_{1})(\vec{x}_{2},h_{2})=(\vec{x}_{1}+h_{1}\vec{x}_{2},h_{1}h_{2}).
\end{equation*}
which can be verified by matrix multiplication using (\ref{eq:matrep-ex-aff-gal}). With the help of the above notations, $g\in\mathcal{G}^{m}_{\hbox{\tiny{aff}}}$ can also be expressed in the following block form
\begin{equation*}
g=\begin{bmatrix}h&\vec{x}\\\vec{0}^{T}&1\end{bmatrix}.
\end{equation*}
Now, let us denote by $\mathfrak{g}$ and $\mathfrak{h}$, the lie algebras of the lie groups $\mathcal{G}^{m}_{\hbox{\tiny{aff}}}$ and $H$ respectively, where the dimension of $\mathfrak{g}$ is six while that of $\mathfrak{h}$ is three.

Now, we assume that $d\mu_{G}$ and $d\mu_r$ are the left and right invariant Haar measures for $\mathcal{G}^{m}_{\hbox{\tiny{aff}}}$ respectively with $\Delta_G$ and $\Delta_H$ being the corresponding modular functions. Then \cite{semidirect}
\begin{equation*}
d\mu_{G}(\vec{x},h)=\Delta_{G}(\vec{x},h)d\mu_{r}(\vec{x},h)=\frac{\Delta_{H}(h)}{|\hbox{det}h|}d\mu_{r}(\vec{x},h)
\end{equation*}
But in section \ref{subsec:haar-msr-ex-aff-gal}, we have already found the following measures and the corresponding modular function
\begin{eqnarray*}
d\mu_{G}(\vec{x},h)&=&e^{-4\sigma+\tau}dv\;da\;db\;d\theta\;d\sigma\;d\tau\\
d\mu_{r}(\vec{x},h)&=&dv\;da\;db\;d\theta\;d\sigma\;d\tau\\
\Delta_{G}(\vec{x},h)&=&e^{-4\sigma+\tau}
\end{eqnarray*}
Also, $|\hbox{det}h|=e^{3\sigma}$. And therefore, we get
\begin{equation}
\Delta_{H}(h)=e^{-\sigma+\tau}.
\end{equation}
Now, we proceed to find the generators for the group $\mathcal{G}^{m}_{\hbox{\tiny{aff}}}$ explcitly and subsequently find a generic group element of $\mathfrak{g}$. Following are the six generators $D_{s},K,D_{T},X,T,\Theta$ corresponding to $\sigma,v,\tau,a,b\; \hbox{and}\;\theta$ respectively,
\begin{eqnarray*}
&&D_{s}=\begin{bmatrix}1&0&0&0\\0&0&0&0\\0&0&2&0\\0&0&0&0\end{bmatrix}, \quad K=\begin{bmatrix}0&1&0&0\\0&0&0&0\\m&0&0&0\\0&0&0&0\end{bmatrix}, \quad D_{T}=\begin{bmatrix}0&0&0&0\\0&1&0&0\\0&0&-1&0\\0&0&0&0\end{bmatrix}\\
&&X=\begin{bmatrix}0&0&0&1\\0&0&0&0\\0&0&0&0\\0&0&0&0\end{bmatrix}, \quad T=\begin{bmatrix}0&0&0&0\\0&0&0&1\\0&0&0&0\\0&0&0&0\end{bmatrix}, \quad \Theta=\begin{bmatrix}0&0&0&0\\0&0&0&0\\0&0&0&1\\0&0&0&0\end{bmatrix}.
\end{eqnarray*}
The commutation relations between the above generators are listed below
\begin{eqnarray}\label{eq:comm-rel-ex-aff-gal}
&&[K,D_{T}]=K,\;[D_{s},D_{T}]=0,\;[K,X]=m\Theta,\;[X,T]=0,\;[K,T]=X\nonumber\\
&&[\Theta,T]=0,\;[\Theta,K]=0,\;[\Theta,X]=0,\;[\Theta,D_{T}]=\Theta,\;[\Theta,D_{s}]=-2\Theta\\
&&[K,D_{s}]=-K,\;[T,D_{T}]=-T,\;[X,D_{s}]=-X,\;[X,D_{T}]=0,\;[T,D_{s}]=0\nonumber.
\end{eqnarray}
Now a generic algebra element $Y$ can be written as
\begin{equation}\label{eq:algbr-elmnt-ex-aff-gal}
Y=x_{1}D_{s}+x_{2}K+x_{3}D_{T}+x_{4}X+x_{5}T+x_{6}\Theta.
\end{equation}
In matrix notation,
\begin{equation}\label{eq:mtrx-algbr-elmnt-ex-aff-gal}
Y=\begin{bmatrix}x_{1}&x_{2}&0&x_{4}\\0&x_{3}&0&x_{5}\\mx_{2}&0&2x_{1}-x_{3}&x_{6}\\0&0&0&0\end{bmatrix}.
\end{equation}
$Y\in\mathfrak{g}$ can conveniently be written as
\begin{equation*}
Y=\begin{bmatrix}X_{q}&\vec{x_p}\\\vec{0}^{T}&0\end{bmatrix},
\end{equation*}
where we let
\begin{equation}\label{eq:sbalgbr-elmnt-ex-aff-gal}
X_{q}=\begin{bmatrix}x_{1}&x_{2}&0\\0&x_{3}&0\\
mx_{2}&0&2x_{1}-x_{3}\end{bmatrix},
\end{equation}
and
\begin{equation*}
\vec{x_p}=\begin{pmatrix}x_4\\x_5\\x_6\end{pmatrix}.
\end{equation*}
We also let
\begin{equation*}
\vec{x_q}=\begin{pmatrix}x_1\\x_2\\x_3\end{pmatrix},
\end{equation*}
so that the six dimensional column vector $\begin{pmatrix}\vec{x_q}\\\vec{x_p}\end{pmatrix}$ represents a generic algebra element. How a generic group element $(\vec{x},h)$ acts on such a six dimensional vector is encoded in the so-called adjoint action matrix while the action of a group element on a dual algebra element (in this case six component row vector) is encapsulated in the coadjoint action matrix. Now we will find the coadjoint action matrix for the group $\mathcal{G}^{m}_{\hbox{\tiny{aff}}}$ explicitly.
The inverse group element is given by
\begin{equation}\label{eq:invrs-grp-elmnt-ex-aff-gal}
g^{-1}=\begin{bmatrix}e^{-\sigma}&-ve^{-\sigma}&0&e^{-\sigma}(vb-a)\\0&e^{-\tau}&0&-be^{-\tau}\\-mve^{\tau-2\sigma}&\frac{1}{2}mv^{2}e^{\tau-2\sigma}&e^{\tau-2\sigma}&e^{\tau-2\sigma}(-\theta+mva-\frac{1}{2}mv^{2}b)\\0&0&0&1\end{bmatrix}
\end{equation}
The adjoint action of the underlying group on a generic lie algebra element is defined as
\begin{equation*}
Ad_{g}Y=gYg^{-1}.
\end{equation*}
And hence
\begin{eqnarray*}
Ad_{g^{-1}}Y&=&g^{-1}Yg\nonumber\\
&=&\begin{bmatrix}x_{1}^{\prime}&x_{2}^{\prime}&0&x_{4}^{\prime}\\0&x_{3}^{\prime}&0&x_{5}^{\prime}\\mx_{2}^{\prime}&0&2x_{1}^{\prime}-x_{3}^{\prime}&x_{6}^{\prime}\\0&0&0&0\end{bmatrix},
\end{eqnarray*}
with
\begin{eqnarray*}
&&x_{1}^{\prime}=x_{1},\; x_{2}^{\prime}=ve^{\tau-\sigma}x_{1}+e^{\tau-\sigma}x_{2}-ve^{\tau-\sigma}x_{3},\; x_{3}^{\prime}=x_{3},\\
&&x_{4}^{\prime}=e^{-\sigma}ax_{1}+e^{-\sigma}bx_{2}-e^{-\sigma}vbx_{3}+e^{-\sigma}x_{4}-ve^{-\sigma}x_{5},\; x_{5}^{\prime}=be^{-\tau}x_{3}+e^{-\tau}x_{5},\\
&&x_{6}^{\prime}=e^{\tau-2\sigma}(2\theta-mva)x_{1}+e^{\tau-2\sigma}(ma-mvb)x_{2}+e^{\tau-2\sigma}(\frac{1}{2}mv^{2}b-\theta)x_{3}\\
&&-mve^{\tau-2\sigma}x_{4}+\frac{1}{2}mv^{2}e^{\tau-2\sigma}x_{5}+e^{\tau-2\sigma}x_{6}.
\end{eqnarray*}
We, therefore, obtain
\begin{equation*}
\scalefont{0.75}\left[\begin{matrix}x_{1}^{\prime}\\x_{2}^{\prime}\\x_{3}^{\prime}\\x_{4}^{\prime}\\x_{5}^{\prime}\\x_{6}^{\prime}\end{matrix}\right]
=\scalefont{0.75}{\left[\,\begin{matrix}1&0&0&0&0&0\\ve^{\tau-\sigma}&e^{\tau-\sigma}&-ve^{\tau-\sigma}&0&0&0\\0&0&1&0&0&0\\ae^{-\sigma}&be^{-\sigma}&-vbe^{-\sigma}&e^{-\sigma}&-ve^{-\sigma}&0\\0&0&be^{-\tau}&0&e^{-\tau}&0\\e^{\tau-2\sigma}(2\theta-mva)&\mkern5mue^{\tau-2\sigma}(ma-mvb)&\mkern5mue^{\tau-2\sigma}(\frac{1}{2}mv^{2}b-\theta)&\mkern5mu-mve^{\tau-2\sigma}&\mkern5mu\frac{1}{2}mv^{2}e^{\tau-2\sigma}&\mkern5mue^{\tau-2\sigma}\end{matrix}\,\right]
\left[\begin{matrix}x_1\\x_2\\x_3\\x_4\\x_5\\x_6\end{matrix}\right]}.
\end{equation*}
So, the coadjoint action matrix for the (1+1) dimensional extended affine Galilei group is given by the following six by six matrix
$$\scalefont{0.75}\left[\,\begin{matrix}1&0&0&0&0&0\\ve^{\tau-\sigma}&e^{\tau-\sigma}&-ve^{\tau-\sigma}&0&0&0\\0&0&1&0&0&0\\ae^{-\sigma}&be^{-\sigma}&-vbe^{-\sigma}&e^{-\sigma}&-ve^{-\sigma}&0\\0&0&be^{-\tau}&0&e^{-\tau}&0\\e^{\tau-2\sigma}(2\theta-mva)&\mkern4mue^{\tau-2\sigma}(ma-mvb)&\mkern4mue^{\tau-2\sigma}(\frac{1}{2}mv^{2}b-\theta)&\mkern4mu-mve^{\tau-2\sigma}&\mkern4mu\frac{1}{2}mv^{2}e^{\tau-2\sigma}&\mkern4mue^{\tau-2\sigma}\end{matrix}\,\right].$$
Now, let us have a closer look at the relevant coadjoint orbits. We already know that the coadjoint orbits are just the cotangent bundles on the dual orbits. So in our case, the non degenerate coadjoint orbits are given by $\mathcal{O}_{1}^*,\mathcal{O}_{2}^{*}, \mathcal{O}_{3}^{*}$, and $\mathcal{O}_{4}^{*}$, where
\begin{eqnarray}\label{eq:coadorb-ex-aff-gal}
\mathcal{O}_{1}^{*}&=&T^{*}\hat{\mathcal{O}_{1}}\nonumber\\
\mathcal{O}_{2}^{*}&=&T^{*}\hat{\mathcal{O}_{2}}\nonumber\\
\mathcal{O}_{3}^{*}&=&T^{*}\hat{\mathcal{O}_{3}}\\
\mathcal{O}_{4}^{*}&=&T^{*}\hat{\mathcal{O}_{4}}\nonumber.
\end{eqnarray}
And the details about $\hat{\mathcal{O}_{1}},\hat{\mathcal{O}_{2}},\hat{\mathcal{O}_{3}},\hat{\mathcal{O}_{4}}$ are outlined in Table \ref{tab:tablefirst}.

Now, let us have a look at how the coadjoint orbits in (\ref{eq:coadorb-ex-aff-gal}) follow directly from the above coadjoint action matrix.
Let the group act on a dual algebra element $(0,0,0,0,k_{1},k_{2})$ via the coadjoint action matrix to give another element of the dual algebra which basically lies in one of the coadjoint orbits. Here it is assumed that $(k_1,k_2)\in\mathbb{R}^{2}\setminus (0,0)$. The resulting element $r$ of a coadjoint orbit is given by
\begin{eqnarray*}
r=[e^{\tau-2\sigma}(2\theta-mva)k_{2},\;e^{\tau-2\sigma}(ma-mvb)k_{2},\;k_{1}be^{-\tau}+e^{\tau-2\sigma}(\frac{1}{2}mv^{2}b-\theta)k_{2},\\
-mvk_{2}e^{\tau-2\sigma},\;k_{1}e^{-\tau}+\frac{1}{2}mv^{2}k_{2}e^{\tau-2\sigma},\;k_{2}e^{\tau-2\sigma}].
\end{eqnarray*}
The last three coordinates of the above vector are basically the coordinates of a point lying in  either of the following three dimensional manifolds $\hat{\mathcal{O}_{1}},\hat{\mathcal{O}_{2}},\hat{\mathcal{O}_{3}}$, and $\hat{\mathcal{O}_{4}}$ depending on the sign of $k_1\;\hbox{and}\;k_{2}$ as we can see from a change of variables
\begin{eqnarray*}
\hat{k}_{1}&=&-mvk_{2}e^{\tau-2\sigma}\\
\hat{k}_{2}&=&k_{2}e^{\tau-2\sigma}\\
\hat{k}_{3}&=&k_{1}e^{-\tau}.
\end{eqnarray*}
Further to this, we also see that as $\theta,b,a,v,\sigma,\tau$ run through $\mathbb{R}$ independently, the first three components of $r$ also run through
$\mathbb{R}$ independently. Denoting them as $k_{1}^{*},k_{2}^{*}$, and $k_{3}^{*}$, respectively, we can hence write $r$ in the following manner
\begin{equation}\label{eq:elmnt-coadorbt-ex-aff-gal}
r=\begin{bmatrix}k_{1}^{*}&k_{2}^{*}&k_{3}^{*}&\hat{k}_1&\frac{(\hat{k}_1)^2}{2m\hat{k}_2}+\hat{k}_3&\hat{k}_2\end{bmatrix}.
\end{equation}
Here, $k_{1}^{*},k_{2}^{*},k_{3}^{*}$ are the vector components relating to the fibre part (cotangent space) of the cotangent bundle while $\hat{k}_1,\hat{k}_2,\hat{k}_3$ are those corresponding to the base manifold (the dual orbit in question). As the sign of $k_1$ and that of $k_2$ determine the fact which coadjoint orbit we are in, we would like to denote the underlying non degenerate coadjoint orbits by $\mathcal{O}_{k_{1},k_{2}}^{*}$. For example, we can take a fixed pair $(k_1,k_2)\in \mathbb{R}^{2}\setminus(0,0)$ such that $k_1,k_2>0$ and act the codjoint action matrix on an algebra element $(0,0,0,0,k_1,k_2)$ to get the coadjoint orbit $\mathcal{O}_{1}^{*}$ as each of the group parameters $\theta,b,a,v,\sigma$ and $\tau$ vary in $\mathbb{R}$. We can write the above fact as $\mathcal{O}^{*}_{\vec{K}_{j}^{T}}=T^{*}\hat{\mathcal{O}}_{\vec{K}_{j}^{T}}$ where $\vec{K}_{j}^{T}=(0,k_1,k_2)$. We can conveniently choose an ordered pair $(1,1)$ for the value of $(k_1,k_2)$ to generate the coadjoint orbit $\mathcal{O}_{1}^{*}$, i.e. $\mathcal{O}_{1}^{*}=\mathcal{O}^{*}_{\vec{K}_{1}^{T}}=T^{*}\hat{\mathcal{O}}_{\vec{K}_{1}^{T}}$, where $\vec{K}_{1}^{T}=(0,1,1)$. We can go on to find the other coadjoint orbits by suitably choosing an ordered pair for the value of $(k_1,k_2)$. Following is the table describing how we obtain different coadjoint orbits due to different signs of the non-zero components of the element $\begin{bmatrix}0&0&0&0&k_1&k_2\end{bmatrix}$ lying in the dual algebra along with the corresponding representative vectors in $\hat{\mathbb{R}}^{3}$. The coadjoint orbits in the following table are also expressed in terms of the dual orbits described in Table \ref{tab:tablefirst}.

\begin{table}[h]
\caption{Classification of the coadjoint orbits $\mathcal{O}_{\vec{K}_{j}^T}^{*}$ depending on the signs of the components of the vector $\vec{K}_{j}^{T}=(0,k_1,k_2)$.}
\label{tab:tablesecond}
\centering
\begin{tabular}{c|c|c|l}
\hline
$\mathcal{O}_{\vec{K}_j^T}^{*}$ & $k_1$ & $k_2$& representative vector $\vec{K}_{j}^{T}$\\
\hline
\hline
$\mathcal{O}_{\vec{K}_{1}^{T}}^{*}:=\mathcal{O}_{1}^{*}=T^{*}\hat{\mathcal{O}}_{1}$ & $>0$ & $>0$ & $\vec{K}_{1}^{T}=(0,1,1)$\\ % inserting body of the table
$\mathcal{O}_{\vec{K}_{2}^{T}}^{*}:=\mathcal{O}_{2}^{*}=T^{*}\hat{\mathcal{O}}_{2}$ & $<0$ & $>0$ &$\vec{K}_{2}^{T}= (0,-1,1)$ \\
$\mathcal{O}_{\vec{K}_{3}^{T}}^{*}:=\mathcal{O}_{3}^{*}=T^{*}\hat{\mathcal{O}}_{3}$ & $<0$ & $<0$ &$\vec{K}_{3}^{T}= (0,-1,-1)$ \\
$\mathcal{O}_{\vec{K}_{4}^{T}}^{*}:=\mathcal{O}_{4}^{*}=T^{*}\hat{\mathcal{O}}_{4}$ & $>0$ & $<0$ &$\vec{K}_{4}^{T}= (0,1,-1)$ \\
\hline
\end{tabular}
\end{table}
\subsection{Necessary ingredients to cook up the Wigner function for (1+1) dimensional extended affine Galilei group}\label{subsec:tools-wigfnc-ex-aff-gal}
We have alreday noted that $\hat{\mathcal{O}}_{\vec{K}_{1}^{T}},\hat{\mathcal{O}}_{\vec{K}_{2}^{T}},\hat{\mathcal{O}}_{\vec{K}_{3}^{T}}$ and $\hat{\mathcal{O}}_{\vec{K}_{4}^{T}}$ are the only four non degenerate orbits for $\mathcal{G}^{m}_{\hbox{\tiny{aff}}}$ with $\cup_{j=1}^{4}\hat{\mathcal{O}}_{\vec{K}_{j}^{T}}$ being dense in $\hat{\mathbb{R}}^3$. And also, $\cup_{j=1}^{4}T^{*}\hat{\mathcal{O}}_{\vec{K}_{j}^T}$ is dense in $\mathbb{R}^6$. In addition to these, the orbits are open free. To be precise, each of the above four dual orbits is an open free H-orbit. We will find an unitarily inequivalent irreducible and square integrable representation due to each such open free orbit. These representations exhaust unitary irreducible representations for the underlying group exactly. And the quasi-regular representation of $\mathcal{G}^{m}_{\hbox{\tiny{aff}}}=\mathbb{R}^{3}\rtimes H$ turns out to be just a direct sum of these four irreducible representations. We speak about quasi-regular representation beacause the underlying Hilbert space is no longer $L^{2}(\mathcal{G}^{m}_{\hbox{\tiny{aff}}},d\mu_{G}(\vec{x},h))$, rather it is $\mathfrak{H}=L^{2}(\mathcal{G}^{m}_{\hbox{\tiny{aff}}}/H\simeq\mathbb{R}^{3},d\theta\;db\;da)$. It is convenient to work in the Fourier transformed space $\hat{\mathfrak{H}}=L^{2}(\hat{\mathbb{R}}^{3},dq\;dE\;dp)$. And the quasi-regular representations in this Fourier-transformed space $\hat{\mathfrak{H}}$ is unitarily equivalent to those defined on the Hilbert space $\mathfrak{H}$. To be precise, the unitary operators $\hat{U}(\vec{x},h)$ acts on the square integrable functions living in the Hilbert space $\hat{\mathfrak{H}}=L^{2}(\hat{\mathbb{R}}^{3},dq\;dE\;dp)$ in the following manner \cite{semidirect}
\begin{eqnarray}\label{eq:uir-ex-aff-gal}
\lefteqn{(\hat{U}(\vec{x},h)\hat{f})(\begin{bmatrix}p&E&q\end{bmatrix})}\nonumber\\
&&=e^{\frac{3\sigma}{2}}e^{i(q\theta+Eb+pa)}\hat{f}\left(\begin{bmatrix}p&E&q\end{bmatrix}\begin{bmatrix}e^{\sigma}&ve^{\tau}&0\\0&e^{\tau}&0\\mve^{\sigma}&\frac{1}{2}mv^{2}e^{\tau}&e^{2\sigma-\tau}\end{bmatrix}\right)\nonumber\\
&&=e^{\frac{3\sigma}{2}}e^{i(q\theta+Eb+pa)}\hat{f}(\begin{bmatrix}e^{\sigma}(p+qmv)&e^{\tau}(pv+E+\frac{1}{2}qmv^{2})e^{\tau}&qe^{2\sigma-\tau}\end{bmatrix}).\nonumber\\
&&\*
\end{eqnarray}
If we set $\hat{\mathfrak{H}}_{j}=L^{2}(\hat{\mathcal{O}}_{\vec{K}_{j}^{T}},dq\;dE\;dp)$, we see that each of these spaces is an invariant subspace of $\hat{U}$. And $\hat{U}_{j}$, the restriction of $\hat{U}$ to the Hilbert space $\hat{\mathfrak{H}_{j}}$ is irreducible. In other words, we have the following direct sum decomposition of $\hat{U}$
\begin{equation}\label{eq:drctsum-decom-uir-ex-aff-gal}
\hat{U}(\vec{x},h)=\oplus_{j=1}^{4}\hat{U}_{j}(\vec{x},h).
\end{equation}
And $\hat{\mathfrak{H}}$, the representation space of $\hat{U}(\vec{x},h)$ decomposes in the following way
\begin{equation}\label{eq:repspc-decom-ex-aff-gal}
\hat{\mathfrak{H}}=\oplus_{1}^{4}\hat{\mathfrak{H}_{j}}.
\end{equation}
Now, that we are done with the business of representations, we move onto computing the Duflo-Moore operator in question. The Duflo-Moore operator pertaining to the open free orbit $\hat{\mathcal{O}}_{\vec{K}_{j}^{T}}$ is defined to be \cite{semidirect}
\begin{equation}\label{eq:duflo-moore-gnrl-expr}
(C_{j}\hat{f})(\vec{k}^{T})=(2\pi)^{\frac{n}{2}}[c_{j}(\vec{k}^{T})]^{\frac{1}{2}}\hat{f}(\vec{k}^{T})
\end{equation}
where $\hat{f}\in L^{2}(\hat{\mathcal{O}}_{\vec{K}_{j}^{T}},d\vec{k}^{T})$ and $c_{j}:\hat{\mathcal{O}}_{\vec{K}_{j}^{T}}\rightarrow\mathbb{R}^{+}$ is a positive Lebesgue measurable function. Let us compute this measurable function explicitly for $\hat{\mathcal{O}}_{\vec{K}_{1}^{T}}$. It is to be noted that this orbit is basically the dual orbit outlined as $\hat{\mathcal{O}}_{1}$ in Table \ref{tab:tablefirst}. We would have had the measurable functions $c$ to be constant, if $\mathcal{G}^{m}_{\hbox{\tiny{aff}}}$ were unimodular. But we will see now that because of the non unimodularity of the (1+1) dimensional extended affine Galilei group we have the function $c_{1}$ to be unbounded above. We first take an arbitrary element $\begin{bmatrix}0&1&1\end{bmatrix}\in\hat{\mathcal{O}}_{\vec{K}_{1}^{T}}(\hbox{because}\;1^{2}-\frac{0^2}{2(m)(1)}>0)$ and let the $3\times 3$ matrix $h$ introduced in section \ref{subsec:coadmtrx-ex-aff-gal} act on this element
\begin{eqnarray}\label{eq:act-subgrp-ex-aff-gal}
\lefteqn{\begin{bmatrix}0&1&1\end{bmatrix}\begin{bmatrix}e^{\sigma}&ve^{\tau}&0\\0&e^{\tau}&0\\mve^{\sigma}&\frac{1}{2}mv^{2}e^{\tau}&e^{2\sigma-\tau}\end{bmatrix}}\nonumber\\
&&=\begin{bmatrix}mve^{\sigma}&e^{\tau}+\frac{1}{2}mv^{2}e^{\tau}&e^{2\sigma-\tau}\end{bmatrix}
\end{eqnarray}
Let us further introduce the following change of variables
\begin{eqnarray}\label{eq:chng-of-var-sub-ex-aff-gal}
mve^{\sigma}&=&k_{1}^{\prime}\nonumber\\
e^{2\sigma-\tau}&=&k_{2}^{\prime}\\
e^{\tau}&=&k_{3}^{\prime}\nonumber.
\end{eqnarray}
With the above change of variables the right side of (\ref{eq:act-subgrp-ex-aff-gal}) takes the form $\begin{bmatrix}k_{1}^{\prime}&k_{3}^{\prime}+\frac{(k_{1}^{\prime})^2}{2mk_{2}^{\prime}}&k_{2}^{\prime}\end{bmatrix}$. As $k_{2}^{\prime}$ and $k_{3}^{\prime }$ are always positive, the vector represented by the last matrix definitely lives in $\hat{\mathcal{O}}_{\vec{K}_{1}^{T}}$. We, therefore, constructed a homeomorphism from $H$, the underlying closed subspace of $GL(3,\mathbb{R})$ to the dual orbit $\hat{\mathcal{O}}_{\vec{K}_{1}^{T}}$. The corresponding lebesgue measures have the following transformation
\begin{equation}\label{eq:lebmeas-tranfm-sub-ex-aff-gal}
dk_{1}^{\prime}\;dk_{2}^{\prime}\;dk_{3}^{\prime}=2me^{3\sigma}dv\;d\sigma\;d\tau.
\end{equation}
Next, we transfer the left invariant Haar measure $d\mu_{H}$ from $GL(3,\mathbb{R})$ to $\hat{\mathcal{O}}_{\vec{K}_{1}^{T}}$ under the above mentioned homeomorphism.

Now, the left invariant Haar measure on $H$ can be computed to be \cite{coherent}
\begin{equation}\label{eq:lft-haarmsr-sub-ex-aff-gal}
d\mu_{H}(h)=e^{-\sigma+\tau}dv\;d\sigma\;d\tau.
\end{equation}
Therefore, we have the following result
\begin{eqnarray}\label{lft-haarmeas-hom-eqn-ex-aff-gal}
d\mu_{H}(h)&=&e^{-\sigma+\tau}dv\;d\sigma\;d\tau\nonumber\\
&=&\frac{e^{-4\sigma+\tau}}{2m}dk_{1}^{\prime}\;dk_{2}^{\prime}\;dk_{3}^{\prime}\nonumber\\
&=&\frac{1}{2m|k_{2}^{\prime}|^{2}|k_{3}^{\prime}|}dk_{1}^{\prime}\;dk_{2}^{\prime}\;dk_{3}^{\prime}.
\end{eqnarray}
From which follows the expression for the function $c_{1}$
\begin{equation}\label{eq:fnc-duflo-moore-ex-aff-gal}
c_{1}(\begin{bmatrix}k_{1}^{\prime}&k_{3}^{\prime}+\frac{(k_{1}^{\prime})^2}{2mk_{2}^{\prime}}&k_{2}^{\prime}\end{bmatrix})=\frac{1}{2m|k_{2}^{\prime}|^{2}|k_{3}^{\prime}|}.
\end{equation}
Now the Duflo-Moore operator $C_{1}$ corresponding to the open free orbit $\hat{\mathcal{O}}_{\vec{K}_{1}^{T}}$ immediately follows from (\ref{eq:duflo-moore-gnrl-expr})
\begin{equation}
(C_{1}\hat{f})(\begin{bmatrix}k_{1}^{\prime}&k_{3}^{\prime}+\frac{(k_{1}^{\prime})^2}{2mk_{2}^{\prime}}&k_{2}^{\prime}\end{bmatrix})=\frac{(2\pi)^{\frac{3}{2}}}{(2m)^{\frac{1}{2}}|k_{2}^{\prime}||k_{3}^{\prime}|^{\frac{1}{2}}}\hat{f}(\begin{bmatrix}k_{1}^{\prime}&k_{3}^{\prime}+\frac{(k_{1}^{\prime})^2}{2mk_{2}^{\prime}}&k_{2}^{\prime}\end{bmatrix}).
\end{equation}

Now, we want to find the adjoint representation of $\mathfrak{h}$, the lie algebra of $H$. A generic element $X_q\in\mathfrak{h}$ is given by (\ref{eq:sbalgbr-elmnt-ex-aff-gal}). The generators $K, D_{s},D_{T}$ as mentioned in Section \ref{subsec:coadmtrx-ex-aff-gal} form a basis for $\mathfrak{h}$. The corresponding commutation relations along with adjoint representations for the bases are given by
\begin{equation}
[K,K]= 0,\;\;[K,D_s]= -K,\;\;[K,D_T]= K,
\end{equation}
leading to
\begin{equation}\label{eq:adrep-ex-aff-gal-boost}
\hbox{ad}K=\begin{bmatrix}0&0&0\\-1&0&0\\1&0&0\end{bmatrix}.
\end{equation}
Also,
\begin{equation}
[D_s,K]=K,\;\;[D_s,D_s]=0,\;\;[D_s,D_T]=0,
\end{equation}
giving
\begin{equation}\label{eq:adrep-ex-aff-gal-spcdil}
\hbox{ad}D_{s}=\begin{bmatrix}1&0&0\\0&0&0\\0&0&0\end{bmatrix}.
\end{equation}
Again,
\begin{equation}
[D_T,K]=-K,\;\;[D_T,D_s]=0,\;\;[D_T,D_T]=0,
\end{equation}
which in turn gives
\begin{equation} \label{eq:adrep-ex-aff-gal-tmedil}
\hbox{ad}D_T=\begin{bmatrix}-1&0&0\\0&0&0\\0&0&0\end{bmatrix}.
\end{equation}
Using (\ref{eq:adrep-ex-aff-gal-boost}), (\ref{eq:adrep-ex-aff-gal-spcdil}), and (\ref{eq:adrep-ex-aff-gal-tmedil}), we can express $\hbox{ad}\;\frac{X_q}{2}$, given a generic algebra element $X_q\in\mathfrak{h}$ with $X_q=x_{2}K+x_{1}D_s+x_{3}D_T$, as
\begin{eqnarray}\label{eq:adrep-subalgbr-ex-aff-gal}
\hbox{ad}\frac{X_q}{2}&=&\frac{x_2}{2}\hbox{ad}K+\frac{x_1}{2}\hbox{ad}D_{s}+\frac{x_3}{2}\hbox{ad}D_{T}\nonumber\\
&=&\begin{bmatrix}\frac{x_{1}-x_{3}}{2}&0&0\\-\frac{x_{2}}{2}&0&0\\\frac{x_{2}}{2}&0&0\end{bmatrix}.
\end{eqnarray}
Now, given an $n\times n$ matrix $A$, we define for notational convenience, $\hbox{sinch}A$ as \cite{plancherel}
\begin{equation}\label{eq:def-sinch}
\hbox{sinch}A=\mathbb{I}_{n}+\frac{1}{3!}A^{2}+\frac{1}{5!}A^{4}+\frac{1}{7!}A^{6}+\dots
\end{equation}
Using (\ref{eq:adrep-subalgbr-ex-aff-gal}) and (\ref{eq:def-sinch}), we immediately obtain
\begin{equation}\label{eq:sinch-ad}
\hbox{sinch}\;(\hbox{ad}\;\frac{X_q}{2})=\begin{bmatrix}\hbox{sinch}\;(\frac{x_{1}-x_{3}}{2})&0&0\\-\frac{x_{2}}{x_{1}-x_{3}}\hbox{sinch}\;(\frac{x_{1}-x_{3}}{2})+\frac{x_{2}}{x_{1}-x_{3}}&1&0\\\frac{x_{2}}{x_{1}-x_{3}}\hbox{sinch}\;(\frac{x_{1}-x_{3}}{2})-\frac{x_{2}}{x_{1}-x_{3}}&0&1\end{bmatrix}.
\end{equation}
Therefore we have,
\begin{equation}\label{eq:det-ad-sinch}
\hbox{det}\;(\hbox{sinch}\;\hbox{ad}\;\frac{X_q}{2})=\hbox{sinch}\;(\frac{x_{1}-x_{3}}{2}).
\end{equation}
We also compute $\hbox{sinch}\;\frac{X_q}{2}$ and $\frac{1}{\hbox{sinch}\;\frac{X_q}{2}}$, which are as follows

\begin{eqnarray*}\label{eq:mat-sinch}
\lefteqn{\hbox{sinch}\left(\frac{X_q}{2}\right)}\\
&=\scalefont{0.65}\left[\,\begin{matrix}\hbox{sinch}(\frac{x_{1}}{2})&\frac{x_{2}}{x_{1}-x_{3}}[\hbox{sinch}(\frac{x_{1}}{2})-\hbox{sinch}(\frac{x_{3}}{2})]&0\\0&\hbox{sinch}(\frac{x_{3}}{2})&0\\\frac{mx_{2}}{x_{1}-x_{3}}[\hbox{sinch}(x_{1}-\frac{x_{3}}{2})-\hbox{sinch}\frac{x_{1}}{2}]&\mkern5mu\frac{mx_{2}^{2}}{2(x_{1}-x_{3})^2}[\hbox{sinch}(\frac{x_{3}}{2})+\hbox{sinch}(x_{1}-\frac{x_{3}}{2})-2\hbox{sinch}(\frac{x_{1}}{2})]&\mkern5mu\hbox{sinch}(x_{1}-\frac{x_{3}}{2})\end{matrix}\,\right].
\end{eqnarray*}
And
\begin{eqnarray}\label{eq:mat-inv-sinch}
\lefteqn{\frac{1}{\hbox{sinch}\left(\frac{X_q}{2}\right)}}\nonumber\\
&=\scalefont{0.7}\left[\,\begin{matrix}\frac{1}{\hbox{sinch}(\frac{x_{1}}{2})}&\frac{x_{2}}{x_{1}-x_{3}}[\frac{1}{\hbox{sinch}(\frac{x_{1}}{2})}-\frac{1}{\hbox{sinch}(\frac{x_{3}}{2})}]&0\\0&\frac{1}{\hbox{sinch}(\frac{x_{3}}{2})}&0\\\frac{mx_{2}}{x_{1}-x_{3}}[\frac{1}{\hbox{sinch}(x_{1}-\frac{x_{3}}{2})}-\frac{1}{\hbox{sinch}(\frac{x_{1}}{2})}]&\mkern6mu\frac{mx_{2}^{2}}{2(x_{1}-x_{3})^2}[\frac{1}{\hbox{sinch}(\frac{x_{3}}{2})}+\frac{1}{\hbox{sinch}(x_{1}-\frac{x_{3}}{2})}-\frac{2}{\hbox{sinch}(\frac{x_{1}}{2})}]&\mkern6mu\frac{1}{\hbox{sinch}(x_{1}-\frac{x_{3}}{2})}\end{matrix}\,\right].\nonumber\\
\end{eqnarray}
A generic lie algebra element $X_q$ of $\mathfrak{h}$ was given by
\begin{equation*}
X_q=\begin{bmatrix}x_{1}&x_{2}&0\\0&x_{3}&0\\mx_{2}&0&2x_{1}-x_{3}\end{bmatrix}.
\end{equation*}
The entries of $X_q$ are all expressed in terms of $x_{1}, x_{2}$ and $x_{3}$. Now we denote the domain of integration by $\mathcal{D}$ where $\mathcal{D}=\{(x_{1},x_{2},x_{3})\in\mathbb{R}^{3}|x_{1}\neq x_{3}\}$. The reason we exclude the points given by $x_{1}=x_{3}$ is that the nonzero offdiagonal entries of the matrix $\hbox{sinch}\;\frac{X_q}{2}$ and those of $\frac{1}{\hbox{sinch}\;(\frac{X_q}{2})}$ all blow up at that point as can easily be verified from (\ref{eq:mat-sinch}) and (\ref{eq:mat-inv-sinch}). Also if we take $X_q\in\mathfrak{h}$ such that $(x_{1},x_{2},x_{3})\in\mathcal{D}$, then the exponential map taking the Lie algebra elements to the underlying group manifold is definitely a bijection onto a dense set of the Lie group $H$.

\subsection{Wigner function for the (1+1)-extended affine Galilei group and the domain for the corresponding function}\label{subsec:compt-wigfnc-ex-aff-gal}
Now, that we have all the essential ingredients, to compute the Wigner function, at our disposal, we go ahead and do it. We will focus on the open free orbit $\hat{\mathcal{O}}_{\vec{K}_{1}^{T}}$. We assume both $k_1$ and $k_2$ to be equal to 1 in (\ref{eq:elmnt-coadorbt-ex-aff-gal}) as we are considering the cotangent bundle on the open free orbit $\hat{\mathcal{O}}_{\vec{K}_{1}^{T}}$.
Finally, suppressing the hats in (\ref{eq:elmnt-coadorbt-ex-aff-gal}), we find that a point on the corresponding six dimensional coadjoint orbit $\mathcal{O}_{1}^{*}$ has coordinates $(k_{1}^{*},k_{2}^{*},k_{3}^{*},k_{1},k_{3}+\frac{(k_{1})^{2}}{2mk_{2}},k_{2})$ where $k_{1}^{*},k_{2}^{*},k_{3}^{*}$ and $k_{1}$ vary freely in $\mathbb{R}$ while $k_{2}$ and $k_{3}$ can assume values only from the positive real axis ($\mathbb{R}^{+}$). The first three coordinates $(k_{1}^{*},k_{2}^{*},k_{3}^{*})$ correspond to the cotangent space  of the underlying cotangent bundle. On the other hand, the last three coordinates correspond to the base manifold of the cotangent bundle, i.e. the open free orbit in question. At this point, we denote the vectors $\begin{bmatrix}k_{1}^{*}&k_{2}^{*}&k_{3}^{*}\end{bmatrix}$ and $\begin{bmatrix}k_{1}&k_{3}+\frac{k_{1}^{2}}{2mk_{2}}&k_{2}\end{bmatrix}$ as $\vec{\gamma}_{q}^{T}$ and $\vec{\gamma}_{p}^{T}$ respectively.

Now if we denote the Hilbert space of Hilbert-Schmidt operators on\\
$\mathfrak{H}=L^{2}(\hat{\mathcal{O}}_{\vec{K}_{1}^{T}},\frac{1}{2m|k_{2}|^{2}|k_{3}|}dk_{1}\;dk_{2}\;dk_{3})$ as $\mathcal{B}_{2}(\mathfrak{H})$, then the Wigner function due to the corresponding coadjoint orbit is essentially a map
$$W:\mathcal{B}_{2}(\mathfrak{H})\rightarrow L^{2}(\mathcal{O}_{1}^{*},\frac{dk_{1}^{*}\;dk_{2}^{*}\;dk_{3}^{*}\;dk_{1}\;
dk_{2}\;dk_{3}}{2m|k_{2}|^{2}|k_{3}|}) .$$ Since the orbit under study is an open free one, the Wigner function due to the corresponding coadjoint orbit is given by the following formula \cite{semidirect}
\begin{eqnarray}\label{eq:gnrl-expr-wigfnc-semdrct}
W(\hat{\phi},\hat{\psi}|(\vec{\gamma}_{q}^{T},\vec{\gamma}_{p}^{T}))&=&\int_{N_{0q}}d\vec{x}_{q}e^{-i\vec{\gamma}_{q}^{T}\vec{x}_{q}}\overline{\hat{\psi}\left(\vec{\gamma}_{p}^{T}\frac{e^{\frac{X_q}{2}}}{\hbox{sinch}\frac{X_q}{2}}\right)}\hat{\phi}\left(\vec{\gamma}_{p}^{T}\frac{e^{-\frac{X_q}{2}}}{\hbox{sinch}\frac{X_q}{2}}\right)\nonumber\\
&\times& c\left(\vec{\gamma}_{p}^{T}\frac{1}{\hbox{sinch}\frac{X_q}{2}}\right)^{-\frac{1}{2}}c(\vec{\gamma}_{p}^{T})^{-\frac{1}{2}}\left|\frac{\hbox{det}\;(\hbox{sinch}\;\hbox{ad}\frac{X_q}{2})}{\hbox{det}\;(\hbox{sinch}\;\frac{X_q}{2})}\right|^{\frac{1}{2}}
\end{eqnarray}
Now, we apply (\ref{eq:fnc-duflo-moore-ex-aff-gal}), (\ref{eq:sinch-ad}), (\ref{eq:det-ad-sinch}), (\ref{eq:mat-sinch}) and (\ref{eq:mat-inv-sinch}) to compute the following
\begin{eqnarray}\label{eq:fctr-contrbt-wigfnc}
\lefteqn{c\left(\vec{\gamma}_{p}^{T}\frac{1}{\hbox{sinch}\frac{X_q}{2}}\right)^{-\frac{1}{2}}c(\vec{\gamma}_{p}^{T})^{-\frac{1}{2}}\left|\frac{\hbox{det}\;(\hbox{sinch}\;\hbox{ad}\frac{X_q}{2})}{\hbox{det}\;(\hbox{sinch}\;\frac{X_q}{2})}\right|^{\frac{1}{2}}}\nonumber\\
&&=\frac{2m|k_{2}|^{2}}{\hbox{sinch}\;(x_{1}-\frac{x_{3}}{2})}\left|\frac{k_{3}(k_{3}+r)}{\hbox{sinch}\;(\frac{x_3}{2})}-\frac{k_{3}r\;\hbox{sinch}\;(x_{1}-\frac{x_{3}}{2})}{\hbox{sinch}^{2}(\frac{x_{1}}{2})}\right|^{\frac{1}{2}}\nonumber\\
&&\times\left[\frac{\hbox{sinch}\;(\frac{x_{1}-x_{3}}{2})}{\hbox{sinch}\;(\frac{x_{1}}{2})\;\hbox{sinch}\;(\frac{x_3}{2})\;\hbox{sinch}\;(x_{1}-\frac{x_3}{2})}\right]^{\frac{1}{2}}
\end{eqnarray}
where $r=\frac{1}{2mk_{2}}\left(k_{1}-\frac{mk_{2}x_{2}}{x_{1}-x_{3}}\right)^{2}$.

Now, using (\ref{eq:fctr-contrbt-wigfnc}) into (\ref{eq:gnrl-expr-wigfnc-semdrct}), we compute the Wigner function corresponding to one of the coadjoint orbits $\mathcal{O}_{1}^*$ for (1+1) dimensional extended affine Galilei group explicitly
\begin{eqnarray}\label{eq:explct-frm-wigfnc-ex-aff-gal}
\lefteqn{W_{1}(\hat{\phi},\hat{\psi}|k_{1}^{*},k_{2}^{*},k_{3}^{*},k_{1},k_{3}+\frac{k_{1}^{2}}{2mk_{2}},k_{2})}\nonumber\\
&=2m|k_{2}|^{2}\int_{\mathcal{D}}dx_{1}\;dx_{2}\;dx_{3}\;e^{-i(x_{1}k_{1}^{*}+x_{2}k_{2}^{*}+x_{3}k_{3}^{*})}\overline{\hat{\psi}\left(\vec{\gamma}_{p}^{T}\frac{e^{\frac{X_q}{2}}}{\hbox{sinch}\frac{X_q}{2}}\right)}\hat{\phi}\left(\vec{\gamma}_{p}^{T}\frac{e^{-\frac{X_q}{2}}}{\hbox{sinch}\frac{X_q}{2}}\right)\nonumber\\
&\times\left[\frac{\hbox{sinch}(\frac{x_{1}-x_{3}}{2})}{\hbox{sinch}(\frac{x_{1}}{2})\;\hbox{sinch}(\frac{x_3}{2})\;\hbox{sinch}^{3}(x_{1}-\frac{x_3}{2})}\right]^{\frac{1}{2}}\left|\frac{k_{3}(k_{3}+r)}{\hbox{sinch}(\frac{x_3}{2})}-\frac{k_{3}r\;\hbox{sinch}\;(x_{1}-\frac{x_{3}}{2})}{\hbox{sinch}^{2}(\frac{x_{1}}{2})}\right|^{\frac{1}{2}}\nonumber\\
\*
\end{eqnarray}
In (\ref{eq:explct-frm-wigfnc-ex-aff-gal}) $\vec{\gamma}_{p}^{T}=\begin{bmatrix}k_{1}&k_{3}+\frac{k_{1}^{2}}{2mk_{2}}&k_{2}\end{bmatrix}$, where $k_{1}\in\mathbb{R}, k_{2}>0$ and $k_{3}>0$.
Also,
\begin{equation*} e^{X_q}=\begin{bmatrix}e^{x_{1}}&\frac{x_{2}(e^{x_1}-e^{x_3})}{x_{1}-x_{3}}&0\\0&e^{x_3}&0\\\frac{mx_{2}[e^{(2x_{1}-x_{3})}-e^{x_1}]}{x_{1}-x_{3}}&\frac{mx_{2}^{2}[e^{x_{3}}-2e^{x_1}+e^{(2x_{1}-x_{3})}]}{2(x_{1}-x_{3})^{2}}&e^{2x_{1}-x_{3}}\end{bmatrix}.
\end{equation*}
As already mentioned, $r=\frac{1}{2mk_{2}}(k_{1}-\frac{mk_{2}x_{2}}{x_{1}-x_{3}})^{2}$.
And $\frac{1}{\hbox{sinch}\;(\frac{X_q}{2})}$ is given by (\ref{eq:mat-inv-sinch}).

Proceeding in the same manner, we can also find the Wigner functions corresponding to the other three coadjoint obits for $\mathcal{G}^{m}_{\hbox{\tiny{aff}}}$. Now we will discuss the domain of all four Wigner functions corresponding to various coadjoint orbits of the underlying group.

If we have a look at the most general expression of Wigner function $W_{\lambda}$ given by (\ref{eq:gnrl-expr-wigfnc-semdrct}), we immediately see that whether or not this function is supported on the corresponding coadjoint orbit $\mathcal{O}_{\lambda}^{*}$ is entirely determined by the fact if the argument of $\hat{\phi}$ and that of $\hat{\psi}$, i.e. $\vec{\gamma}_{p}^{T}\frac{e^{\frac{X_q}{2}}}{\hbox{sinch}\;\frac{X_q}{2}}$ and $\vec{\gamma}_{p}^{T}\frac{e^{-\frac{X_q}{2}}}{\hbox{sinch}\;\frac{X_q}{2}}$ always stay inside the dual orbit $\hat{\mathcal{O}}_{\lambda}$ or not, which in turn implies that we have to ensure that the vector $\vec{\gamma}_{p}^{T}\in\hat{\mathcal{O}}_{\lambda}$ remains stable under the $``\hbox{sinch}''$ map to have the Wigner function supported on its coadjoint orbit.

Now, the polynomial function $\Delta$ as introduced in \cite{semidirect} reduces, for the (1+1) dimensional affine Galilei group case, to
\begin{eqnarray}\label{eq:pol-func-ex-aff-gal}
\Delta(\begin{pmatrix}p&E&q\end{pmatrix})=\hbox{det}\begin{bmatrix}{\begin{pmatrix}p&E&q\end{pmatrix}}\;K\\\;{\begin{pmatrix}p&E&q\end{pmatrix}}\;D_{s}\\\;\;{\begin{pmatrix}p&E&q\end{pmatrix}}\;D_{T}\end{bmatrix}.
\end{eqnarray}
Here, $\begin{pmatrix}p&E&q\end{pmatrix}\in \mathbb{R}^{3}$, where the dual orbits are all embedded. And as we already know,
\begin{equation}
K=\begin{bmatrix}0&1&0\\0&0&0\\m&0&0\end{bmatrix},\;D_{s}=\begin{bmatrix}1&0&0\\0&0&0\\0&0&2\end{bmatrix},\;D_{T}=\begin{bmatrix}0&0&0\\0&1&0\\0&0&-1\end{bmatrix}.
\end{equation}
(\ref{eq:pol-func-ex-aff-gal}) now reduces to
\begin{eqnarray}\label{eq:explct-expr-pol-fnc-orbt-shp}
\Delta(\begin{pmatrix}p&E&q\end{pmatrix})&=&\hbox{det}\begin{bmatrix}qm&p&0\\p&0&2q\\0&E&-q\end{bmatrix}\nonumber\\
&=&q(p^{2}-2mqE)
\end{eqnarray}
So, in terms of this new polynomial function $\Delta$, we can construct a table for the non degenerate dual orbits, using Table \ref{tab:tablefirst}, which is as follows
\begin{table}[thb]
\renewcommand{\arraystretch}{1}
\caption{Classification of orbits of $\mathcal{G}^{m}_{\hbox{aff}}$ in $\mathbb{R}^3$ under $H=\mathcal{V}\rtimes\mathbb{R}^2$ }
\label{tab:tablethird}
\centering
\begin{tabular}{c|c|c|c}
\hline
Orbits & $q$ & $\Delta(\begin{pmatrix}p&E&q\end{pmatrix})$ & $E-p\;\hbox{relation}$\\
\hline
\hline
$\hat{\mathcal{O}}_{1}$ & $>0$ & $<0$ & $E>\frac{p^{2}}{2mq}$ \\ % inserting body of the table
$\hat{\mathcal{O}}_{2}$ & $>0$ & $>0$ & $E<\frac{p^{2}}{2mq}$ \\
$\hat{\mathcal{O}}_{3}$ & $<0$ & $<0$ & $E<\frac{p^{2}}{2mq}$ \\
$\hat{\mathcal{O}}_{4}$ & $<0$ & $>0$ & $E>\frac{p^{2}}{2mq}$\\
\hline
\end{tabular}
\end{table}

We can easily see from the above table that the sign of $\Delta$ changes as we move back and forth between $\hat{\mathcal{O}}_{1}$ and $\hat{\mathcal{O}}_{2}$. The same is true for $\hat{\mathcal{O}}_{3}$ and $\hat{\mathcal{O}}_{4}$.
Now, we take an arbitrary element $(p_0,\;E_0,\;q_0)$
from one of the nondegenerate dual orbits and then we act $\frac{1}{\hbox{sinch}\;\frac{X_q}{2}}$ on it to obtain the following vector in $\mathbb{R}^{3}$,
\begin{eqnarray}
\lefteqn{v=\biggl(\frac{p_0}{\hbox{sinch}(\frac{x_1}{2})}+\frac{mx_{2}q_{0}}{x_{1}-x_{3}}\left[\frac{1}{\hbox{sinch}(x_{1}-\frac{x_3}{2})}-\frac{1}{\hbox{sinch}(\frac{x_1}{2})}\right],}\nonumber\\
&&\frac{p_{0}x_{2}}{x_{1}-x_{3}}\left[\frac{1}{\hbox{sinch}(\frac{x_1}{2})}-\frac{1}{\hbox{sinch}(\frac{x_3}{2})}\right]+\frac{E_0}{\hbox{sinch}(\frac{x_3}{2})}+\frac{mx_{2}^{2}q_{0}}{2(x_{1}-x_{3})^{2}}\nonumber\\\
&&\times\left[\frac{1}{\hbox{sinch}(x_{1}-\frac{x_3}{2})}+\frac{1}{\hbox{sinch}(\frac{x_3}{2})}-\frac{2}{\hbox{sinch}\;(\frac{x_1}{2})}\right],\frac{q_0}{\hbox{sinch}(x_{1}-\frac{x_3}{2})}\biggr)\nonumber
\end{eqnarray}
We observe that the sign of $q_0$ is invariant under the above transformation. In other words, if we start with a point lying in $\hat{\mathcal{O}}_{1}$ it can leak into $\hat{\mathcal{O}}_{2}$ at best. It can never leak down to $\hat{\mathcal{O}}_{3}$ or to $\hat{\mathcal{O}}_{4}$ through the $q=0$ plane. Similarly, if we start with a point in $\hat{\mathcal{O}}_{3}$ we can end up with a point in $\hat{\mathcal{O}}_{4}$ under the action of the ``$\frac{1}{\hbox{sinch}}$'' map. But the point can never go across the $q=0$ plane to reach either to $\hat{\mathcal{O}}_{1}$ or to $\hat{\mathcal{O}}_{2}$.
Next, we compute the polynomial function $\Delta$ given by (\ref{eq:explct-expr-pol-fnc-orbt-shp}) at the point given by the vector $v$ explicitly.
\begin{eqnarray}\label{eq:pol-fnc-atapoint}
\lefteqn{\Delta(v)}\nonumber\\
&&=\hbox{det}\scalefont{0.75}{\left[\,\begin{matrix}mq_{0}D&p_{0}L+\frac{mx_{2}q_{0}}{x_{1}-x_{3}}(D-L)&0\\p_{0}L+\frac{mx_{2}q_{0}}{x_{1}-x_{3}}(D-L)&0&2q_{0}D\\0&\mkern2mu{\frac{p_{0}x_{2}}{x_{1}-x_{3}}}(L-T)+E_{0}T+\frac{mx_{2}^{2}{q_{0}}}{2(x_{1}-x_{3})^{2}}(D+T-2L)&\mkern2mu-q_{0}D\end{matrix}\,\right]}
\end{eqnarray}
where we have assumed the following
\begin{eqnarray}\label{eq:inv-sinch-at-algbr}
\frac{1}{\hbox{sinch}\;(\frac{x_1}{2})}&=&L\nonumber\\
\frac{1}{\hbox{sinch}\;(\frac{x_3}{2})}&=&T\\
\frac{1}{\hbox{sinch}\;(x_{1}-\frac{x_{3}}{2})}&=&D\nonumber,
\end{eqnarray}
Computing the determinant given by the right side of (\ref{eq:pol-fnc-atapoint}) we have,
\begin{eqnarray}\label{eq:pol-fnc-atapoint-evltd}
\Delta(v)=m^{2}q_{0}^{3}D(L^{2}-DT)\frac{x_{2}^{2}}{(x_{1}-x_{3})^{2}}-2mp_{0}q_{0}^{2}D(L^{2}-DT)\frac{x_2}{x_{1}-x_{3}}\nonumber\\
+q_{0}D(p_{0}^{2}L^{2}-2mq_{0}E_{0}DT).
\end{eqnarray}
Setting $\Delta(v)$ to be zero and letting $\frac{x_{2}}{x_{1}-x_{3}}=s$ we get
\begin{eqnarray}\label{eq:sprt-dtrmng-eq}
\lefteqn{m^{2}q_{0}^{3}(L^{2}-DT)\left(s^{2}-\frac{2p_{0}}{mq_{0}}s\right)+q_{0}(p_{0}^{2}L^{2}-2mq_{0}E_{0}DT)=0}\nonumber\\
&\Rightarrow& m^{2}q_{0}^{3}(L^{2}-DT)(s-\frac{p_{0}}{q_{0}})^{2}=q_{0}p_{0}^{2}m(L^{2}-DT)-q_{0}(p_{0}^{2}L^{2}-2mq_{0}E_{0}DT)\nonumber\\
&\Rightarrow& \left(s-\frac{p_{0}}{q_{0}}\right)^{2}=\frac{DT(2mq_{0}E_{0}-p_{0}^{2})}{m^{2}q_{0}^{2}(L^{2}-DT)}.
\end{eqnarray}
It can be verified that $L^{2}-DT>0$. Also, $DT>0$. Therefore, what (\ref{eq:sprt-dtrmng-eq}) tells us is that a point $(p_{0},\;E_{0},\;q_{0})$ in one of the non degenerate dual orbits can only be unstable under the action of ``$\frac{1}{\hbox{sinch}}$'' map iff $p_{0}^{2}-2mq_{0}E_{0}<0$. But
\begin{eqnarray}\label{eq:spprt-coad-orbt}
&&p_{0}^{2}-2mq_{0}E_{0}<0\nonumber\\
&&\Rightarrow (p_{0},\;E_{0},\;q_{0})\in \hat{\mathcal{O}}_{1}\; \hbox{or}\;(p_{0},\;E_{0},\;q_{0})\in \hat{\mathcal{O}}_{4},
\end{eqnarray}
which can easily be seen from Table \ref{tab:tablethird}. We also find that the points in $\hat{\mathcal{O}}_{2}$ and those in $\hat{\mathcal{O}}_{3}$ are all stable under the ``$\frac{1}{\hbox{sinch}}$'' map, i.e. they do not leave the corresponding orbits under the action of that map.

The Wigner function corresponding to the coadjoint orbit $\mathcal{O}_{\lambda}^{*}$, as a function of $\vec{\gamma}_{q}^{T}\in\hat{\mathbb{R}}^{3}$ can be thought of as the Fourier transform of a function $F(X_q)$
\begin{equation}\label{eq:fourier-wigfnc}
W_{\vec{\omega}_{0}^{T}}(\hat{\phi},\hat{\psi}|\mathcal{O}_{\lambda}^{*})=\int_{\mathcal{D}}\;d\vec{x}_{q}\;e^{-i\vec{\gamma}_{q}^{T}\vec{x_q}}F(X_q),
\end{equation}
where
\begin{eqnarray}\label{eq:fourier-trfm}
F(X_q)&=&\overline{\hat{\psi}\left(\vec{\omega}_{0}^{T}\frac{e^{\frac{X_q}{2}}}{\hbox{sinch}\frac{X_q}{2}}\right)}\hat{\phi}\left(\vec{\omega}_{0}^{T}\frac{e^{-\frac{X_q}{2}}}{\hbox{sinch}\frac{X_q}{2}}\right)\nonumber\\
&\times& c\left(\vec{\omega}_{0}^{T}\frac{1}{\hbox{sinch}\frac{X_q}{2}}\right)^{-\frac{1}{2}}c(\vec{\omega}_{0}^{T})^{-\frac{1}{2}}\left|\frac{\hbox{det}\;(\hbox{sinch}\;\hbox{ad}\frac{X_q}{2})}{\hbox{det}\;(\hbox{sinch}\;\frac{X_q}{2})}\right|^{\frac{1}{2}}.
\end{eqnarray}
Also, $\hat{\phi},\hat{\psi}\in L^{2}(\hat{\mathcal{O}}_{\lambda})$ and $\vec{\omega}_{0}^{T}$ is any point from one of the four disjoint nondegenerate dual orbits.

Now, for $\lambda=2$, if $W_{\vec{\omega}_{0}^{T}}(\hat{\phi},\hat{\psi}|\mathcal{O}_{2}^{*})$ were supported on $\mathcal{O}_{2}^{*}$, then $F(X_q)$ would have been identically zero if we chose $\vec{\omega}_{0}^{T}\notin\hat{\mathcal{O}}_{2}$. But we have already seen that if we take $\vec{\omega}_{0}^{T}\in\hat{\mathcal{O}}_{3}\;\hbox{or}\;\hat{\mathcal{O}}_{4}$, $\vec{\omega}_{0}^{T}\frac{1}{\hbox{sinch}\frac{X_q}{2}}$ can never be in $\hat{\mathcal{O}}_{2}$. On the other hand, (\ref{eq:spprt-coad-orbt}) tells us that if $\vec{\omega}_{0}^{T}\in\hat{\mathcal{O}}_{1}$ we can end up with $\vec{\omega}_{0}^{T}\frac{1}{\hbox{sinch}\frac{X_q}{2}}\in\hat{\mathcal{O}}_{2}$. In other words, $F(X_q)$ can assume  nonzero values when $\vec{\omega}_{0}^{T}\in\hat{\mathcal{O}}_{1}$. Therefore the Wigner function corresponding to $\mathcal{O}_{2}^{*}$ will have its support spread on both the coadjoint orbits $\mathcal{O}_{1}^{*}$ and $\mathcal{O}_{2}^{*}$.

Now, we consider $\lambda=1$ in equation (45). So, $\hat{\phi},\hat{\psi}\in L^{2}(\hat{\mathcal{O}}_{1})$. Now, the question that we are going to address is whether we can have $\vec{\omega}_{0}^{T}\notin\hat{\mathcal{O}}_{1}$ such that $\vec{\omega}_{0}^{T}\frac{1}{\hbox{sinch}\frac{X_q}{2}}\in\hat{\mathcal{O}}_{1}$.
It is obvious from our previous discussion that we can not have such a point in $\hat{\mathbb{R}}^{3}$. Again, for an element $\vec{\omega}_{0}^{T}\in\hat{\mathcal{O}}_{1}$ and $\vec{\omega}_{0}^{T}\frac{1}{\hbox{sinch}\frac{X_q}{2}}\notin\hat{\mathcal{O}}_{1}$ we have $F(X_q)$ to be identically zero as $\hat{\phi},\hat{\psi}\in L^{2}(\hat{\mathcal{O}}_{1})$ by assumption. Therefore, we find the support of $W_{\vec{\omega}_{0}^{T}}(\hat{\phi},\hat{\psi}|\mathcal{O}_{1}^{*})$ always lying inside $\mathcal{O}_{1}^{*}$.

Using exactly the same arguments we find that the Wigner function  $W_{\vec{\omega}_{0}^{T}}(\hat{\phi},\hat{\psi}|\mathcal{O}_{4}^{*})$ is supported inside $\mathcal{O}_{4}^{*}$ while the support of  $W_{\vec{\omega}_{0}^{T}}(\hat{\phi},\hat{\psi}|\mathcal{O}_{3}^{*})$ is spread out on both the coadjoint orbits $\mathcal{O}_{3}^{*}$ and $\mathcal{O}_{4}^{*}$.

We, therefore, conclude that the Wigner functions corresponding to the two coadjoint orbits $\mathcal{O}_{1}^{*}$ and $\mathcal{O}_{4}^{*}$ will have their supports concentrated inside $\mathcal{O}_{1}^{*}$ and $\mathcal{O}_{4}^{*}$ respectively. It is to be noted that the zero level sets of the polynomial function $\Delta$ introduced earlier in this section, restricted to $\mathcal{O}_{1}^{*}$ and $\mathcal{O}_{4}^{*}$ are not decomposable into hyperplanes. These two zero-level sets are the two dimensional surfaces in Figure \ref{fig:figsecond} (above and below the plane $q=0$) correspoding to the degenerate orbits $\hat{\mathcal{O}}_{5}$ and $\hat{\mathcal{O}}_{6}$ respectively in Table \ref{tab:tablefirst}. And hence we have verified that the converse of the following theorem due to A. E. Krasowska and S. T. Ali \cite{semidirect} is not true.
\newtheorem{theorem}{Theorem}
\begin{Theo}\cite{semidirect}
Let $G$ be a semi-direct product group $\mathbb{R}^{n}\rtimes H$, such that $H$ acts on $\hat{\mathbb{R}}^{n}$ with open free orbits $\{\hat{\mathcal{O}}_{i}^{m}\}_{i=1}$. If an orbit $\hat{\mathcal{O}}_{i}$ is a dihedral cone (i.e. if the zero level set of the polynomial function $\Delta$, restricted to it, can be decomposed into hyperplanes) then the Wigner function $W_{\hat{\mathcal{O}}_{i}}$ has support concentrated on the corresponding coadjoint orbit $\mathcal{O}_{i}^{*}=\mathbb{R}^{n}\times\hat{\mathbb{O}}_{i}$.
\end{Theo}
The sufficient condition for the Wigner function to be supported on one of its coadjoint orbits is that the corresponding dual orbit be a dihedral cone. However, it is not a necessary condition for the Wigner function to have its support inside one of its coadjoint orbits as we can see from the example of (1+1) dimensional extended affine Galilei group.

\section{Wigner function for the (1+1)-centrally extended Galilei\\ group}\label{sec:wigfnc-ex-gal-grp}
We extend the (1+1)-Galilei group $\mathcal{G}_{0}$ centrally using the canonical exponent $\xi(g,g^{\prime})$ given by
\begin{equation}\label{eq:can-expt}
\xi_{1}(g_{1},g_{2})=m(v_{1}a_{2}+\frac{1}{2}v_{1}^{2}b_{2}),
\end{equation}
where $g\equiv(b_{1},a_{1},v_{1})$ and $g_{2}\equiv(b_{2},a_{2},v_{2})$ are elements of $\mathcal{G}_{0}$.
The centrally extended Galilei group $\mathcal{G}^{m}$ then obeys the following group law
\begin{eqnarray}\label{eq:grp-law-ex-gal-can}
\lefteqn{(\theta_{1},b_{1},a_{1},v_{1})(\theta_{2},b_{2},a_{2},v_{2})}\nonumber\\
&&=(\theta_{1}+\theta_{2}+m[v_{1}a_{2}+\frac{1}{2}v_{1}^{2}b_{2}],b_{1}+b_{2},a_{1}+a_{2}+v_{1}b_{2},v_{1}+v_{2}).
\end{eqnarray}
This group $\mathcal{G}^{m}$ has been called the quantum Galilei group in \cite{affgal}. Our first goal in this section would be to study this quantum Galilei group $\mathcal{G}^{m}$ in detail and subsequently to find its coadjoint orbits. As will turn out later in this section that by using the standard procedures \cite{plancherel}, one fails to compute the correct Wigner function for $\mathcal{G}^{m}$ built on the relevant coadjoint orbits. In order to remedy this problem we consider a new exponent $\xi_{2}$ \cite{levy} of the (1+1)-Galilei group $\mathcal{G}_{0}$ which is equivalent to $\xi_{1}$ and is given by (\ref{eq:can-expt}),

\begin{equation}\label{eq:new-expt}
\xi_{2}(g_1,g_2)=\frac{1}{2}m(-v_{1}v_{2}b_{2}+v_{1}a_{2}-v_{2}a_{1}).
\end{equation}

Next, we will extend $\mathcal{G}_{0}$ centrally using the exponent $\xi_{2}$ given by (\ref{eq:new-expt}) and denote the resulting group as $\mathcal{G}^{m\prime}$. The group composition law for $\mathcal{G}^{m\prime}$ is as follows

\begin{eqnarray}\label{eq:grp-law-ex-gal-new}
\lefteqn{(\theta_1,b_1,a_1,v_1)(\theta_2,b_2,a_2,v_2)}\nonumber\\
&&=(\theta_1+\theta_2+\frac{1}{2}m(-v_{1}v_{2}b_{2}+v_{1}a_{2}-v_{2}a_{1}),b_{1}+b_{2},a_{1}+a_{2}+v_{1}b_{2},v_{1}+v_{2})\nonumber\\
&&\*
\end{eqnarray}
We will then find the coadjoint action matrix for $\mathcal{G}^{m\prime}$ which will turn out to be exactly the same as to be found for the quantum Galilei group $\mathcal{G}^{m}$. In other words, the geometry of the coadjoint orbits remains unchanged. And we will arrive at the correct form of Wigner functions using this nontrivial central extension of the (1+1)-Galilei group as we will explore by the end of this section.
\subsection{Dual orbits and the induced representation for the quantum Galilei group $\mathcal{G}^{m}$}\label{subsec:dual-orb-rep-qntm-gal}
The group element for the (1+1)-centrally extended Galilei group $\mathcal{G}^{m}$ or the quantum Galilei group is found to be the following
\begin{equation} \label{eq:mat-rep-qntm-gal}
g= \begin{bmatrix}
      1&v&0&a\\
      0&1&0&b\\
      mv&\frac{1}{2}mv^2&1&\theta\\
      0&0&0&1
      \end{bmatrix},
\end{equation}
which we here denote as $ (\theta,b,a,v)$. And the corresponding group multiplication is given by (\ref{eq:grp-law-ex-gal-can}).

The inverse group element is given by
\begin{equation*}
(\theta,b,a,v)^{-1}=(-\theta-\frac{1}{2}mv^2b+mva,-b,vb-a,-v).
\end{equation*}
In matrix notation
\begin{equation*}
(\theta,b,a,v)^{-1}=\begin{bmatrix}1&-v&0&vb-a\\0&1&0&-b\\-mv&\frac{1}{2}mv^2&1&-\theta-\frac{1}{2}mv^2b+mva\\0&0&0&1\end{bmatrix}.
\end{equation*}
The (1+1) dimensional extended Galilei group $\mathcal{G}^{m}$ or the quantum Galilei group can be viewed as $(\Theta\times\mathcal{T}\times\mathcal{S})\rtimes\mathcal{V}$. The action of a pure Galilian boost $v\in\mathcal{V}$ on the abelian subgroup $(\theta,b,a)\in(\Theta\times\mathcal{T}\times\mathcal{S})$ is computed in the following way
\begin{eqnarray*}
\lefteqn{(\theta_1,b_1,a_1,v_1)(\theta_2,b_2,a_2,v_2)}\\
&&=((\theta_1,b_1,a_1)+v_1(\theta_2,b_2,a_2),v_1+v_2).
\end{eqnarray*}
But according to the group multiplication law, given at the beginning the last expression should equal
$(\theta_1+\theta_2+\frac{1}{2}mv_{1}^2b_2+mv_1a_2,b_1+b_2,a_1+a_2+v_1b_2,v_1+v_2)$,
from which it follows that
\begin{equation*}
v(\theta,b,a)=(\theta+\frac{1}{2}mv^2b+mva,b,a+vb)
\end{equation*}
Let us assume that the dual of the abelian subgroup $\Theta\times\mathcal{T}\times\mathcal{S}$ is parametrized by $\gamma$, $E$ and $p$ where $E,p\in\mathbb{R}$ and $\gamma\in\mathbb{R}\setminus\{0\}$. The $\gamma=0$ case will be handled separately. Now the dual pairing reads
\begin{equation*}
\chi_{\gamma,E,p}(\theta,b,a)=\exp[i(\gamma\theta+Eb+pa)]
\end{equation*}
The dual action of $v$ on the character group can be defined by the following relation
\begin{equation*}
\langle (v)\chi_{\gamma,E,p};(\theta,b,a)\rangle=\langle\chi_{\gamma,E,p};v^{-1}(\theta,b,a)\rangle
\end{equation*}
But it is seen that $v^{-1}=-v$.
From which it follows immediately that
\begin{align*}
v^{-1}(\theta,b,a)&=(-v)(\theta,b,a)\\
&=(\theta+\frac{1}{2}mv^2b-mva,b,a-vb).
\end{align*}
Now,
\begin{eqnarray*}
\lefteqn{\chi_{\gamma,E,p}(\theta+\frac{1}{2}mv^2b-mva,b,a-vb)}\\
&&=\exp i[\gamma\theta+\frac{m\gamma v^2}{2}b-m\gamma va+Eb+p(a-vb)]
\end{eqnarray*}
And, therefore it follows that
\begin {align*}
\chi_{\gamma^{\prime},E^{\prime},p^{\prime}}(\theta,b,a)&=\mathrm{e}^{i[\gamma^{\prime}\theta+E^{\prime}b+p^{\prime}a]}\\
&=\mathrm{e}^{i[\gamma\theta+(\frac{m\gamma v^2}{2}+E-pv)b+(p-m\gamma v)a]}
\end{align*}
So, under the dual group action the variables parameterizing the character group transforms according to the following equations
\begin{eqnarray}\label{eq:dual-crdnt-trnfm-quant-gal}
\gamma^{\prime}&=&\gamma,\nonumber\\
           E^{\prime}&=&\frac{m\gamma v^2}{2}+E-pv,\\
p^{\prime}&=& p-m\gamma v\nonumber.
\end{eqnarray}
We also find $E$ and $p$ to be constrained by an equation which follows from the following computation
\begin{eqnarray*}
p^{\prime 2}&=& p^2-2m\gamma pv+m^2\gamma^2v^2\\
\frac{p^{\prime 2}}{2m\gamma}&=& \frac{p^2}{2m\gamma}-pv+\frac{1}{2}m\gamma v^2.
\end{eqnarray*}
Using the transformation rule for $E$, we immidiately have,
\begin{equation*}
\frac{p^{\prime 2}}{2m\gamma}=\frac{p^2}{2m\gamma}+E^{\prime}-E,
\end{equation*}
which we can rewrite as
\begin{equation*}
E-\frac{p^2}{2m\gamma}=E^{\prime}-\frac{p^{\prime 2}}{2m \gamma}=E_0
\end{equation*}
i.e,
\begin{equation*}
E^\prime=E_0+\frac{p^{\prime 2}}{2m \gamma}
\end{equation*}
So, the dual action on the character group can conveniently be written as
\begin{equation}\label{eq:dual-act-qnt-gal}
(v)\chi_{\gamma,E,p}=\chi_{\gamma^\prime,E^\prime,p^\prime}=\chi_{\gamma,\frac{p^{\prime 2}}{2m\gamma}+E_0,p^\prime}
\end{equation}
For a fixed value of $\gamma$ and that of $E_0$ the orbit is represented by a parabola parallel to the $E^{\prime}p^{\prime}$ plane and perpendicular to the $\gamma$-axis. As $\gamma$ varies over $\mathbb{R}\setminus \{0\}$ the parabola changes its shape continuously. Now, for $\gamma=0$, the dual orbits are computed separately by putting $\gamma$ to be zero in (\ref{eq:dual-crdnt-trnfm-quant-gal}). The corresponding orbits turn out to be simply one dimensional lines parallel to $E^{\prime}$-axis lying in the $\gamma=0$-plane. On the other hand, for $\gamma\neq 0$, the parabolas derived earlier tend to shrink down to lines parallel to $E^\prime$-axis as $\gamma\rightarrow 0$. At the other extreme, the parabolas tend to widen with the increase of $|\gamma|$ and are almost  lines parallel to the $p^\prime$-axis when $|\gamma|$ is large enough. We will just consider the parabolic orbits arising from $\gamma\neq 0$ and $(E,p)\in\mathbb{R}^{2}$ in this work since the contribution of the $\gamma=0$-plane in the representation level will be extremely meager.

Having found the dual orbits for the (1+1)-extended Galilei group or the quantum Galilei group $\mathcal{G}^{m}$ in (1+1)-dimension, we now proceed to find all the unitary irreducible representations of this group using the method suggested by Mackey.

We have already found that each dual orbit $\mathcal{O}_{\gamma,E_0}$ for the underlying Lie group is parametrized by two numbers $\gamma, E_0$. Now, we choose a representative $\chi_{\gamma,E_0,0}$ from each orbit $\mathcal{O}_{\gamma,E_0}$ for distinct ordered pairs of $(E_0,\gamma)\in \mathbb{R}\times\mathbb{R}^*$. Mackey's inducing construction suggests that to each dual orbit there corresponds a unitarily inequivalent irreducible representation. In other words, for each ordered pair $(E_0, \gamma)$, we shall obtain a unitary irreducible representation $U^{E_0, \gamma}$ and for any two different ordered pairs the representations will be unitarily inequivalent.
First, it is evident using (\ref{eq:dual-crdnt-trnfm-quant-gal}) that the stabilizer subgroup of $\mathcal{V}$ which leaves a particular character group element, say $\chi_{\gamma,E,p}$ dual to $(\theta,b,a)\in \Theta \times \mathcal{T}\times S$, stable, is the trivial identity element ($0$) of $\mathcal{V}$.
\begin{equation*}
((v)\chi_{\gamma,E,p}=\chi_{\gamma,E,p})\Longrightarrow v=0,
\end{equation*}
we, therefore, find
\begin{equation*}
\mathcal{O}_{\gamma,E_0}\simeq \mathcal{V}/\{\mbox{Identity element}\}\simeq \hat{\mathbb{R}}.
\end{equation*}
Now, according to the general theory, the irreducible representations of the (1+1) dimensional extended Galilei group $\mathcal{G}^{m}$, i.e. $(\Theta\times \mathcal{T}\times S)\rtimes \mathcal{V}$ can now be obtained from the UIR's $V_{\gamma,E_0}$ of the subgroup $(\Theta \times \mathcal{T}\times S)$, where
\begin{eqnarray}\label{eq:subgrp-uir-indcd}
V_{\gamma,E_0}(\theta,b,a)&=&\chi_{\gamma,E_0,0}(\theta,b,a)\nonumber\\
                          &=& \exp i[\gamma \theta+E_{0}b].
\end{eqnarray}
Because we are looking for UIR's due to a fixed orbit, we keep both $\gamma$ and $E_0$ fixed in (\ref{eq:subgrp-uir-indcd}).
Now,
\begin{equation*}
\mathcal{V}/\{\mbox{Id}\}\simeq \mathcal{G}^{m}/(\Theta \times \mathcal{T}\times S)\simeq \hat{\mathbb{R}}.
\end{equation*}
We define the section $\lambda:\mathcal{G}^{m}/(\Theta\times\mathcal{T}\times S)\simeq \hat{\mathbb{R}}\rightarrow \mathcal{G}^{m}$ by
\begin{equation*}
\lambda(k)=(0,0,0,\frac{k}{m})
\end{equation*}
Therefore we have,
\begin{eqnarray*}
g^{-1}\lambda(k)&=&(-\theta-\frac{1}{2}mv^{2}b+mva,-b,vb-a,-v)(0,0,0,\frac{k}{m})\\
&=& (-\theta-\frac{1}{2}mv^{2}b+mva,-b,vb-a,\frac{k}{m}-v)\\
&=& (0,0,0,\frac{k}{m}-v)(-\theta-\frac{k^{2}b}{2m}+ka,-b,\frac{k}{m}b-a,0),
\end{eqnarray*}
which gives the cocycle $h:\mathcal{G}^{m}\times \hat{\mathbb{R}}\rightarrow\Theta\times\mathcal{T}\times S$ with
\begin{equation*}
h(g^{-1},k)=(-\theta-\frac{k^{2}b}{2m}+ka,-b,\frac{k}{m}b-a,0).
\end{equation*}
Therefore,
\begin{equation*}
h(g^{-1},k)^{-1}=(\theta+\frac{k^{2}b}{2m}-ka,b,a-\frac{k}{m}b,0).
\end{equation*}
Now,
\begin{eqnarray*}
V_{\gamma,E_0}(h(g^{-1},k)^{-1})&=&V_{\gamma,E_{0}}(\theta+\frac{k^{2}b}{2m}-ka,b,a-\frac{k}{m}b,0)\\
&=&\exp i[\gamma(\theta+\frac{k^{2}b}{2m}-ka)+E_{0}b]
\end{eqnarray*}
Therefore, we obtain the representation $U^{\gamma,E_{0}}$ of $\mathcal{G}^{m}$ induced from the UIR $V_{\gamma, E_{0}}$ of the subgroup $\Theta\times \mathcal{T}\times S$. This representation acts on the Hilbert space $L^{2}(\hat{\mathbb{R}},dk)$ in the following way
\begin{equation}\label{eq:uir-quant-gal}
(\hat{U}^{\gamma,E_{0}}(\theta,b,a,v)\hat{\phi})(k)=\exp i[\gamma(\theta+\frac{k^{2}b}{2m}-ka)+E_{0}b]\hat{\phi}(k-mv),
\end{equation}
for all $\hat{\phi}\in L^{2}(\hat{\mathbb{R}},dk)$. We see that each $(E_{0},\gamma)\in \mathbb{R}\times \mathbb{R}^*$ gives rise to a unitary irreducible representation $\hat{U}^{\gamma,E_{0}}$ of $\mathcal{G}^{m}$. Further to this, if we take two distinct points $(E_{0},\gamma)$ and $(E^{\prime}_{0},\gamma^{\prime})$ in the representation space $\mathbb{R}\times \mathbb{R}^{*}$ and label the corresponding UIR's as $\hat{U}^{\gamma,E_{0}}$ and $\hat{U}^{\gamma^{\prime},E^{\prime}_{0}}$ we observe that there does not exist a bounded linear operator $V$ on $L^{2}(\hat{\mathbb{R}},dk)$ such that for all $(\theta,b,a,v)\in\mathcal{G}^{m}$ the following holds
\begin{equation*}
V\hat{U}^{\gamma,E_{0}}(\theta,b,a,v)V^{*}=\hat{U}^{\gamma^{\prime},E^{\prime}_{0}}(\theta,b,a,v)
\end{equation*}
which implies that the UIR's pertaining to different orbits given by (\ref{eq:uir-quant-gal}) are unitarily inequivalent.
\subsection{Coadjoint orbits of $\mathcal{G}^{m}$ and comparison with the dual orbits found from the Mackey construction}\label{subsec:coadorb-quant-gal}
The (1+1) dimensional extended Galilei group $\mathcal{G}^{m}$ or the quantum Galilei group is a real Lie group. Let us find the corresponding Lie algebra. We denote the group generators with $K, X, T,\Theta$ corresponding to the group parameters $v, a, b, \theta$ respectively.
Using the expression for a generic group element in matrix form from (\ref{eq:mat-rep-qntm-gal}) the group generators are found to be the following
\begin{equation*}
K=\begin{bmatrix}0&1&0&0\\0&0&0&0\\m&0&0&0\\0&0&0&0\end{bmatrix},\;X=\begin{bmatrix}0&0&0&1\\0&0&0&0\\0&0&0&0\\0&0&0&0\end{bmatrix},\;T=\begin{bmatrix}0&0&0&0\\0&0&0&1\\0&0&0&0\\0&0&0&0\end{bmatrix},\;\Theta=\begin{bmatrix}0&0&0&0\\0&0&0&0\\0&0&0&1\\0&0&0&0\end{bmatrix}.
\end{equation*}
And the corresponding commutation relations are given by
\begin{eqnarray*}
[K,X]= m\Theta,\;[X,T]= 0,\;[K,T]= X,\;[\Theta,T]=0,\;[\Theta,K]=0,\;[\Theta,X]=0.
\end{eqnarray*}
A general element of the lie algebra is given by
\begin{eqnarray*}
Y&=& a_1K+a_2X+a_3T+a_4\Theta\\
 &=& \begin{bmatrix}0&a_1&0&a_2\\0&0&0&a_3\\m a_1&0&0&a_4\\0&0&0&0\end{bmatrix}.
\end{eqnarray*}
Now, the adjoint action of the lie group on a generic lie algebra element is as follows
\begin{eqnarray*}
Ad_{g}Y&=&gYg^{-1}\\
&=&\begin{bmatrix}0&0&0&-ba_1+a_2+va_3\\0&0&0&a_3\\ma_1&0&0&-ma_1+mva_2+\frac{1}{2}mv^2a_3+a_4\\0&0&0&0\end{bmatrix}.
\end{eqnarray*}
Also, the coadjoint action of the group on the dual algebra is defined by the following relation
\begin{equation}\label{eq:def-coad-act}
\langle Ad_{g}^{\#}(Y^*);Y\rangle=\langle Y^*;Ad_{g^{-1}}(Y)\rangle
\end{equation}
$Ad_{g^{-1}}(Y)$ is found to be
\begin{equation*}
Ad_{g^{-1}}(Y)=\begin{bmatrix}0&a_1&0&ba_1+a_2-va_3\\0&0&0&a_3\\ma_1&0&0&m(a-vb)a_1-mva_2+\frac{1}{2}mv^2a_3+a_4\\0&0&0&0\end{bmatrix}
\end{equation*}
Now, we make the following identifications
\begin{equation*}
Y=\begin{bmatrix}0&a_1&0&a_2\\0&0&0&a_3\\ma_1&0&0&a_4\\0&0&0&0\end{bmatrix}\rightarrow\begin{bmatrix}a_1\\a_2\\a_3\\a_4\end{bmatrix}
\end{equation*}
\begin{equation*}
Ad_{g^{-1}}(Y)=\begin{bmatrix}0&a_{1}^{\prime}&0&a_{2}^{\prime}\\0&0&0&a_{3}^{\prime}\\ma_{1}^{\prime}&0&0&a_{4}^{\prime}\\0&0&0&0\end{bmatrix}\rightarrow\begin{bmatrix}a_{1}^{\prime}\\a_{2}^{\prime}\\a_{3}^{\prime}\\a_{4}^{\prime}\end{bmatrix},
\end{equation*}
where,
\begin{eqnarray*}
&&a_{1}^{\prime}= a_{1},\;a_{2}^{\prime}= ba_1+a_2-va_3,\\
&&a_{3}^{\prime}= a_3,\;a_{4}^{\prime}= m(a-vb)a_1-mva_2+\frac{1}{2}mv^{2}a_3+a_4.
\end{eqnarray*}
And
\begin{equation*}
Y^{*}\rightarrow\begin{bmatrix}a_1&a_2&a_3&a_4\end{bmatrix}.
\end{equation*}
With the above identifications, we have,
\begin{eqnarray*}
\langle Y^{*};Ad_{g^{-1}}(Y)\rangle&=& \begin{bmatrix}a_1&a_2&a_3&a_4\end{bmatrix}\begin{bmatrix}a_{1}^{\prime}\\a_{2}^{\prime}\\a_{3}^{\prime}\\a_{4}^{\prime}\end{bmatrix}\\
                            &=& \begin{bmatrix}a_1&a_2&a_3&a_4\end{bmatrix}\begin{bmatrix}1&0&0&0\\b&1&-v&0\\0&0&1&0\\m(a-vb)&-mv&\frac{1}{2}mv^2&1\end{bmatrix}\begin{bmatrix}a_1\\a_2\\a_3\\a_4\end{bmatrix}\\
                            &=& \langle Ad_{g}^{\#}(Y^*);Y\rangle.
\end{eqnarray*}
Therefore, the coadjoint action of the underlying Lie group on the generic dual algebra element $Y^{*}$ is identified with the following matrix
\begin{equation}\label{eq:coad-act-mat-quant-galgrp}
M(g^{-1})=\begin{bmatrix}1&0&0&0\\b&1&-v&0\\0&0&1&0\\m(a-vb)&-mv&\frac{1}{2}mv^2&1\end{bmatrix}
\end{equation}
Now, we explicitly calculate the coadjoint orbits under the group action on the dual algebra elements $\begin {bmatrix}1&0&0&0\end{bmatrix}$, $\begin{bmatrix}0&1&0&0\end{bmatrix}$, $\begin{bmatrix}0&0&1&0\end{bmatrix}$ and $\begin{bmatrix}0&0&0&1\end{bmatrix}$.
\begin{equation*}
\begin{bmatrix}1&0&0&0\end{bmatrix}\begin{bmatrix}1&0&0&0\\b&1&-v&0\\0&0&1&0\\m(a-vb)&-mv&\frac{1}{2}mv^2&1\end{bmatrix}=\begin{bmatrix}1&0&0&0\end{bmatrix},
\end{equation*}
\begin{equation*}
\begin{bmatrix}0&1&0&0\end{bmatrix}\begin{bmatrix}1&0&0&0\\b&1&-v&0\\0&0&1&0\\m(a-vb)&-mv&\frac{1}{2}mv^2&1\end{bmatrix}=\begin{bmatrix}b&1&-v&0\end{bmatrix},
\end{equation*}
\begin{equation*}
\begin{bmatrix}0&0&1&0\end{bmatrix}\begin{bmatrix}1&0&0&0\\b&1&-v&0\\0&0&1&0\\m(a-vb)&-mv&\frac{1}{2}mv^2&1\end{bmatrix}=\begin{bmatrix}0&0&1&0\end{bmatrix},
\end{equation*}
And
\begin{equation*}
\begin{bmatrix}0&0&0&1\end{bmatrix}\begin{bmatrix}1&0&0&0\\b&1&-v&0\\0&0&1&0\\m(a-vb)&-mv&\frac{1}{2}mv^2&1\end{bmatrix}=\begin{bmatrix}m(a-vb)&-mv&\frac{1}{2}mv^2&1\end{bmatrix}.
\end{equation*}
The first and the third orbit are just points in $\mathbb{R}^4$, while the second one traces a two dimensional plane in $\mathbb{R}^4$ as $b,v$ keep on varying on the real line. The fourth orbit can be regarded as the cotangent bundle on a parabola which is again homeomorphic to $\mathbb{R}^2$. So we are basically obtaining two kinds of orbits, namely, one zero dimensional (point) and the other being two dimensional (plane).
Now, if we take the dual algebra-element $\begin{bmatrix}0&0&k_1&k_2\end{bmatrix}$ and compute the corresponding coadjoint orbits for $k_1,k_2$, each varying on the real line subject to the condition that they are not both zero, we get a dense subspace in $\mathbb{R}^4$.
Let us have a closer look at how this dense subspace looks like.
\begin{eqnarray}\label{eq:geom-coad-orb-quant-gal}
\lefteqn{\begin{bmatrix}0&0&k_1&k_2\end{bmatrix}\begin{bmatrix}1&0&0&0\\b&1&-v&0\\0&0&1&0\\m(a-vb)&-mv&\frac{1}{2}mv^2&1\end{bmatrix}}\nonumber\\
&&=\begin{bmatrix}mk_2(a-vb)&-mvk_2&k_1+\frac{1}{2}mv^2k_2&k_2\end{bmatrix}
\end{eqnarray}
where $a,v,b$ varies independently on the real line and $k_1, k_2$ also varies independently subject to the condition that they are not both zero. Let $\begin {bmatrix}1&0&0&0\end{bmatrix}$, $\begin{bmatrix}0&1&0&0\end{bmatrix}$, $\begin{bmatrix}0&0&1&0\end{bmatrix}$, and $\begin{bmatrix}0&0&0&1\end{bmatrix}$ be the basis vectors generating $\mathbb{R}^4$ and also denote the corresponding orthogonal axes by $X,Y,Z,W$. An arbitrary element in $\mathbb{R}^4$ is denoted as $\begin{bmatrix}x&y&z&w\end{bmatrix}$ where $x,y,z, \mbox{and}\hspace{1mm} w$ coordinatise the components along $X,Y,Z, \mbox{and}\hspace{1mm}W$ axes respectively. With this picture in mind, $\mathbb{R}^4$ can be considered as a stack of parallel $\mathbb{R}^3$- hyperplanes othogonal to $W$-axis. For points in such an $\mathbb{R}^3$- hyperplane $x,y, \mbox{and}\hspace{1mm}z$-coordinates take their values independently on the respective axes while the value of the $w$ coordinate is kept fixed. And, there is a unique $\mathbb{R}^3$-hyperplane which is orthogonal to the $W$-axis and passes through the origin. Any point on this hyperplane is designated by $\begin{bmatrix}x&y&z&0\end{bmatrix}$. For the sake of definiteness, we denote this particular $\mathbb{R}^3$-hyperplane by $\mathbb{R}^{3}_{0}$.

Now, if we put $k_2$ to be zero in (\ref{eq:geom-coad-orb-quant-gal}), we immediately see that the orbit reduces to points consisting of $\begin{bmatrix}0&0&k_1&0\end{bmatrix}$ for $k_1\in\mathbb{R}\smallsetminus \{0\}$ which is just the $Z$-$\mbox{axis}\smallsetminus\{0\}$, abbreviated as $Z^*$-axis. Therefore, in view of (\ref{eq:geom-coad-orb-quant-gal}), the total orbit space due to dual algebra-elements in the form of $\begin{bmatrix}0&0&k_1&k_2\end{bmatrix}$ where $k_1$ and $k_2$ are not both zero (punctured $k_1$-$k_2$ plane), denoted as $\mathcal{O}^{*}_{k_1,k_2}$ turns out to be $[(\mathbb{R}^4\smallsetminus\mathbb{R}^3_0)\cup(Z^*$-$\mbox{axis})]$.

Next, we consider dual-algebra elements having the form of $\begin{bmatrix}0&k_3&0&0\end{bmatrix}$ where $k_3\in\mathbb{R}\smallsetminus\{0\}$ and compute the corresponding coadjoint orbit space due to the elements of this form

\begin{equation}\label{eq:nonzero-scnd-comp-dual-algbr}
\begin{bmatrix}0&k_3&0&0\end{bmatrix}\begin{bmatrix}1&0&0&0\\b&1&-v&0\\0&0&1&0\\m(a-vb)&-mv&\frac{1}{2}mv^2&1\end{bmatrix}=\begin{bmatrix}bk_3&k_3&-vk_3&0\end{bmatrix}
\end{equation}
Equation (\ref{eq:nonzero-scnd-comp-dual-algbr}) determines the fact that the corresponding coadjoint orbit space (the union of all the coadjoint orbits for different nonzero values of $k_3$) denoted as $\mathcal{O}^{*}_{k_3}$ becomes $[\mathbb{R}^{3}_{0}\smallsetminus(X$-$Z\hspace{1mm}\mbox{plane})]$.

Now, we consider dual algebra elements of the form $\begin{bmatrix}k_4&0&k_5&0\end{bmatrix}$, where $k_4$ and $k_5$ are not both zero. The coadjoint action gives
\begin{equation}\label{eq:first-third-nonzero-compt}
\begin{bmatrix}k_4&0&k_5&0\end{bmatrix}\begin{bmatrix}1&0&0&0\\b&1&-v&0\\0&0&1&0\\m(a-vb)&-mv&\frac{1}{2}mv^2&1\end{bmatrix}=\begin{bmatrix}k_4&0&k_5&0\end{bmatrix}
\end{equation}
From (\ref{eq:first-third-nonzero-compt}), we easily see that the total coadjoint orbit space due to elements of the form $\begin{bmatrix}k_4&0&k_5&0\end{bmatrix}$ denoted as $\mathcal{O}^{*}_{k_4,k_5}$,where $k_4$ and $k_5$ are not both zero is found to be just the $[X$-$Z \hspace{1mm}\mbox{plane}\smallsetminus \{0\}]$ which we denote as $(X$-$Z\hspace{1mm}\mbox{plane})^*$.

Finally, the coadjoint orbit due to the dual-algebra element $\begin{bmatrix}0&0&0&0\end{bmatrix}$ denoted as $\mathcal{O}^{*}_{0}$ is just the origin of $\mathbb{R}^4$, i.e, $\begin{bmatrix}0&0&0&0\end{bmatrix}$.
The union of the above coadjoint orbits constitutes $\mathbb{R}^4$:
\begin{eqnarray}\label{eq:total-coad-orb-spc}
\lefteqn{\mathcal{O}^{*}_{k_1,k_2}\cup\mathcal{O}^{*}_{k_3}\cup\mathcal{O}^{*}_{k_4,k_5}\cup\mathcal{O}^{*}_{0}}\nonumber\\
&&=[(\mathbb{R}^4\smallsetminus\mathbb{R}^{3}_{0})\cup(Z^{*}\mbox{-}\mbox{axis})]\cup[\mathbb{R}^{3}_{0}\smallsetminus(X\mbox{-}Z\:\mbox{plane})]\cup(X\mbox{-}Z\:\mbox{plane})^*\cup\{0\}\nonumber\\
&&=\mathbb{R}^4.
\end{eqnarray}

Now, we consider a subset of $\mathcal{O}^{*}_{k_1,k_2}$ given in (\ref{eq:total-coad-orb-spc}), say $\mathcal{O}^{*\prime}_{k_1,k_2}$, where instead of puncturing $k_1$-$k_2$ plane we just throw the $k_1$-axis off the $k_1$-$k_2$ plane. In other words, we are interested in the coadjoint orbits due to dual algebra-elements of the form $\begin{bmatrix}0&0&k_1&k_2\end{bmatrix}$ where $k_2$ is never zero. We can then rewrite (\ref{eq:geom-coad-orb-quant-gal}) as
\begin{eqnarray}\label{eq:throw-one-axis-off}
\lefteqn{\begin{bmatrix}0&0&k_1&k_2\end{bmatrix}\begin{bmatrix}1&0&0&0\\b&1&-v&0\\0&0&1&0\\m(a-vb)&-mv&\frac{1}{2}mv^2&1\end{bmatrix}}\nonumber\\
&&=\begin{bmatrix}mk_2(a-vb)&-mvk_2&k_1+\frac{1}{2}mv^2k_2&k_2\end{bmatrix}\nonumber\\
&&=\begin{bmatrix}mk_2(a-vb)&-mvk_2&k_1+\frac{1}{2mk_2}(-mvk_2)^2&k_2\end{bmatrix},
\end{eqnarray}
where (\ref{eq:throw-one-axis-off}) definitely makes sense because $k_2\neq 0$. Now, we observe that the dual orbits of the underlying Lie group (Mackey orbits) are given by triples of the form $\begin{bmatrix}p^{\prime}&E_0+\frac{p^{\prime 2}}{2m\gamma}&\gamma\end{bmatrix}$ in $p^{\prime}E^{\prime}\gamma$ space ($\mathbb{R}^3$), $\gamma$ being unequal to zero. We have already seen in Section \ref{subsec:dual-orb-rep-qntm-gal} that each $(E_0,\gamma)\in \mathbb{R}\times\mathbb{R}^{*}$ gives rise to a parabolic dual orbit in $p^{\prime}E^{\prime}\gamma$ space, characterized by $\mathcal{O}_{\gamma,E_0}$. The triple $\begin{bmatrix}p^{\prime}&E_0+\frac{p^{\prime 2}}{2m\gamma}&\gamma\end{bmatrix}$ coincides with the last three components of (\ref{eq:throw-one-axis-off}) under the following identification
\begin{eqnarray}\label{eq:idnfc-orb-coadorb}
-mvk_2&\leftrightarrow& p^{\prime}\nonumber\\
k_1&\leftrightarrow& E_0.\\
k_2&\leftrightarrow&\gamma\nonumber
\end{eqnarray}
Equation (\ref{eq:throw-one-axis-off}) represents the cotangent bundle on the family of parabolas given by $$\begin{bmatrix}-mvk_2&k_1+\frac{1}{2mk_2}(-mvk_2)^2&k_2\end{bmatrix},$$ where $v_1,k_1\in\mathbb{R}$ and $k_2\in\mathbb{R}\setminus \{0\}$ and $\begin{bmatrix}-mvk_2&k_1+\frac{1}{2mk_2}(-mvk_2)^2&k_2\end{bmatrix}$represents exactly the Mackey orbits under the identification given in (\ref{eq:idnfc-orb-coadorb}).

We, therefore, obtain the correspondence of dual orbits of (1+1) dimensional Extended Galilei group with its coadjoint orbits which is encapsulated in the following equation
\begin{equation}\label{eq:corr-dual-coadorb}
T^{*}\mathcal{O}_{\gamma,E_0}=\mathcal{O}^{*\prime}_{k_1,k_2}
\end{equation}
\subsection{Invariant measure on $\mathcal{G}^{m}$ and the Kirillov 2-forms on its nontrivial coadjoint orbits }\label{subsec:inv-msr-kirill-frm}
Let a group element $g$ given in (\ref{eq:mat-rep-qntm-gal}) be acted upon by a fixed group element $g_0$ from the left to yield the following
\begin{eqnarray*}
g_{0}g&=&\begin{bmatrix}1&v_0&0&a_0\\0&1&0&b_0\\mv_0&\frac{1}{2}mv_{0}^{2}&1&\theta_{0}\\0&0&0&1\end{bmatrix}\begin{bmatrix}1&v&0&a\\0&1&0&b\\mv&\frac{1}{2}mv^2&1&\theta\\0&0&0&1\end{bmatrix}\\
      &=&\begin{bmatrix}1&v+v_0&0&a+bv_0+a_0\\0&1&0&b+b_0\\m(v+v_0)&mvv_0+\frac{1}{2}m(v^2+v^{2}_{0})&1&\theta+\theta_0+mv_{0}a+\frac{1}{2}mv^{2}_{0}b\\0&0&0&1\end{bmatrix}
\end{eqnarray*}
Under the left action of the group the measure $dv\wedge da\wedge db\wedge d\theta$ transforms as
\begin{eqnarray*}
dv^{\prime}\wedge da^{\prime}\wedge db^{\prime}\wedge d\theta^{\prime}&=&dv\wedge(da+v_{0}db)\wedge db\wedge (d\theta+mv_{0}da+\frac{1}{2}mv^{2}_{0}db)\\
&=&(dv\wedge da+v_{0}dv\wedge db)\wedge db \wedge (d\theta+mv_{0}da+\frac{1}{2}mv^{2}_{0}db)\\
&=&(dv\wedge da \wedge db)\wedge (d\theta+mv_{0}da+\frac{1}{2}mv^{2}_{0}db)\\
&=& dv\wedge da \wedge db \wedge d\theta.
\end{eqnarray*}
Therefore, it follows immediately that $dv \wedge da\wedge db\wedge d\theta$ is a left invariant Haar measure for the underlying Lie group.

Now, we act $g_0$ on $g$ from the right to obtain the following
\begin{equation*}
gg_{0}=\begin{bmatrix}1&v_0+v&0&a+a_0+vb_0\\0&1&0&b+b_0\\m(v+v_0)&mvv_{0}+\frac{1}{2}m(v^2+v^{2}_{0})&1&\theta+\theta_{0}+mva_{0}+\frac{1}{2}mv^{2}b_{0}\\0&0&0&1\end{bmatrix}.
\end{equation*}
Under this right action of the group, the measure $dv \wedge da \wedge db\wedge d\theta$ transforms according to
\begin{eqnarray*}
dv^{\prime}\wedge da^{\prime} \wedge db^{\prime}\wedge d\theta^{\prime}&=& dv\wedge(da+b_{0}dv)\wedge db \wedge(ma_{0}dv+mvb_{0}dv+d\theta)\\
&=&(dv \wedge da\wedge db)\wedge (ma_{0}dv+mvb_{0}dv+d\theta)\\
&=& dv \wedge da \wedge db \wedge d\theta.
\end{eqnarray*}
Therefore, $dv \wedge da\wedge db \wedge d\theta$ tuns out to be both a left and right invariant, i.e. an invariant Haar measure for the (1+1)- dimensional Extended Galilei group.

Now, let us take a fixed dual algebra-element $\begin{bmatrix}0&k_0&0&0\end{bmatrix}$ and find its coadjoint orbits.
\begin{eqnarray*}
\begin{bmatrix}0&k_0&0&0\end{bmatrix}\begin{bmatrix}1&0&0&0\\b&1&-v&0\\0&0&1&0\\m(a-vb)&-mv&\frac{1}{2}mv^2&1\end{bmatrix}&=&\begin{bmatrix}bk_0&k_0&-vk_0&0\end{bmatrix}\\
&=& \begin{bmatrix}a_1&k_0&-a_2&0\end{bmatrix},
\end{eqnarray*}
where $k_0\neq 0$ is fixed, $a,v,b \in \mathbb{R}$ and $a_1=bk_0$ and $a_2=vk_0$. The coadjoint orbit is $\mathbb{R}^2$, parameterized by two independent variables $a_1$ and $a_2$. Now if we take a fixed group element and act it on this coadjoint orbit we have,
\begin{eqnarray*}
\lefteqn{\begin{bmatrix}a_1&k_0&-a_2&0\end{bmatrix}\begin{bmatrix}1&0&0&0\\b_0&1&-v_0&0\\0&0&1&0\\m(a_0-vb_0)&-mv_0&\frac{1}{2}mv^{2}_{0}&1\end{bmatrix}}\\
&& =\begin{bmatrix}a_1+b_{0}k_{0}&k_0&-k_{0}v_{0}-a_2&0\end{bmatrix}\\
&&:= \begin{bmatrix}a_{1}^{\prime}&k_0&-a_{2}^{\prime}&0\end{bmatrix}.
\end{eqnarray*}
We observe that the coadjoint orbit here is stable under the coadjoint action of a fixed group element. Next thing to see is that if we define a two-form on this coadjoint orbit as $ da_1\wedge da_2$ where $a_1$ and $a_2$ have been defined as above, we immediately find it to be invariant under the coadjoint action:
\begin{equation*}
da^{\prime}_{1}\wedge da^{\prime}_{2}=da_{1}\wedge da_{2}.
\end{equation*}
This is the well-known Kirillov 2-form for the coadjoint orbit under study.

Now, we study the same for the other non-trivial coadjoint orbit of the (1+1) dimensional extended Galilei group. The dual algebra element under consideration is now $\begin{bmatrix}0&0&0&k_0\end{bmatrix}$. It's coadjoint orbit is given by
\begin{eqnarray*}
\lefteqn{\begin{bmatrix}0&0&0&k_0\end{bmatrix}\begin{bmatrix}1&0&0&0\\b&1&-v&0\\0&0&1&0\\m(a-vb)&-mv&\frac{1}{2}mv^2&1\end{bmatrix}}\\
&&=\begin{bmatrix}mk_{0}(a-vb)&-mvk_{0}&\frac{1}{2}mv^{2}k_{0}&k_0\end{bmatrix}\\
&&=\begin{bmatrix}a_1&a_2&\frac{a_{2}^{2}}{2mk_0}&k_0\end{bmatrix},
\end{eqnarray*}
where $k_0\neq 0$ is fixed, $a,v,b\in \mathbb{R}$, and $a_1=mk_{0}(a-vb)$ and $a_2=-mvk_0$. Let us find how this two dimensional coadjoint orbit behaves under the coadjoint action of a fixed group element.
\begin{eqnarray*}
\lefteqn{\begin{bmatrix}a_1&a_2&\frac{a_{2}^{2}}{2mk_0}&k_0\end{bmatrix}\begin{bmatrix}1&0&0&0\\b_0&1&-v_0&0\\0&0&1&0\\m(a_0-v_{0}b_{0})&-mv_{0}&\frac{1}{2}mv^{2}_{0}&1\end{bmatrix}}\\
&&=\begin{bmatrix}a_1+b_{0}a_{2}+mk_{0}(a_{0}-v_{0}b_{0})&a_2-k_{0}mv_0&-a_{2}v_{0}+\frac{a^{2}_{2}}{2mk_{0}}+\frac{1}{2}mv^{2}_{0}k_{0}&k_0\end{bmatrix}\\
&&=\begin{bmatrix}a_1+b_{0}a_{2}+mk_{0}(a_{0}-v_{0}b_{0})&a_2-k_{0}mv_{0}&\frac{(a_2-k_{0}mv_{0})^2}{2mk_0}&k_0\end{bmatrix}\\
&&:=\begin{bmatrix}a^{\prime}_{1}&a^{\prime}_{2}&\frac{(a^{\prime}_{2})^2}{2mk_0}&k_0\end{bmatrix}.
\end{eqnarray*}
From the above computation, we observe that this two dimensional coadjoint orbit (cotangent bundle of a parabola) is also stable under the coadjoint action. And, under the aforementioned definition of $a_1, a_2, a_{1}^{\prime}, \mbox{and}\; a_{2}^{\prime}$, it turns out that $da_1\wedge da_2$ is the invariant Kirillov two-form on the coadjoint orbit under study
\begin{eqnarray*}
da_{1}^{\prime}\wedge da_{2}^{\prime}&=&(da_1+b_{0}da_2)\wedge da_2\\
                                &=&da_1\wedge da_2.
\end{eqnarray*}
\subsection{Duflo-Moore operator and Plancherel measure for $\mathcal{G}^{m}$}\label{subsec:duflo-moore-planch-meas-quant-gal}
We start with the following orthogonality condition \cite{plancherel}
\begin{eqnarray}\label{eq:ortho cond}
\lefteqn{\int_{G}\{\int_{\hat{G}}\mbox{tr}([U^{\gamma,E_{0}}(x)^{*}A^{1}(\gamma,E_{0})C_{\gamma,E_{0}}^{-1}])d\nu_{G}(\gamma,E_{0})}\nonumber\\
&\times\int_{\hat{G}}\mbox{tr}([U^{\gamma^{\prime},E^{\prime}_{0}}(x)^{*}A^{2}(\gamma^{\prime},E^{\prime}_{0})C^{-1}_{\gamma^{\prime},E^{\prime}_{0}}])d\nu_{G}(\gamma^{\prime},E^{\prime}_{0})\}d\mu(x)=\langle, A^{1}|A^{2}\rangle_{\mathcal{B}_{2}^{\oplus}}
\end{eqnarray}
where $\mathcal{B}_{2}$ is the underlying Hilbert space of Hilbert-Schmidt operators defined on the representation space (indexed by a certain set of parameters) of the unitary irreducible representations of the given group. And these Hilbert spaces generally vary as we keep varying the corresponding parameters determining the relevant representation spaces.
In the present scenario, we take $A^{1}(\gamma,E_0)=A^{2}(\gamma,E_0)=|\psi_{\gamma,E_0}\rangle\langle\phi_{\gamma,E_0}|$ and the Plancherel measure as $d\nu_{G}(\gamma,E_{0})=\rho(\gamma,E_{0})dE_{0}d\gamma$, so that we have
\begin{eqnarray}\label{eq:ortho-con-presnt-secnario}
\langle A^{1}|A^{1}\rangle_{\mathcal{B}_{2}^{\oplus}}&=&\int_{(E_{0},\gamma)\in\mathbb{R}\times\mathbb{R}^{*}}\mbox{tr}[A^{1}(\gamma,E_{0})^{*}A^{1}(\gamma,E_{0})]\rho(\gamma,E_{0})d\gamma dE_{0}\nonumber\\
&=&\int_{(E_{0},\gamma)\in\mathbb{R}\times\mathbb{R}^{*}}\mbox{tr}[|\phi_{\gamma,E_{0}}\rangle\langle\psi_{\gamma,E_{0}}|\psi_{\gamma,E_{0}}\rangle\langle\phi_{\gamma,E_{0}}|]\rho(\gamma,E_{0})d\gamma dE_{0}\nonumber\\
&=& \int_{(E_{0},\gamma)\in\mathbb{R}\times\mathbb{R}^{*}}\|\psi_{\gamma,E_{0}}\|^{2}\|\phi_{\gamma,E_{0}}\|^{2}\rho(\gamma,E_{0})d\gamma dE_{0}.
\end{eqnarray}
Now, subject to $A^{1}(\gamma,E_0)=A^{2}(\gamma,E_0)=|\psi_{\gamma,E_0}\rangle\langle\phi_{\gamma,E_0}|$ and the fact that the underlying group is unimodular so that the celebrated Duflo-Moore operator $C_{\gamma,E_{0}}$ is just a multiple of the identity operator acting on the Hilbert space $L^{2}(\hat{\mathbb{R}},dk)$, the left side of (\ref{eq:ortho cond}) in the Fourier transformed space reads (from now on we will compute things in mommentum-space representation which is tractable compared to one in configuration space)
\begin{eqnarray*}
\lefteqn{\frac{1}{N^2}\int_{\mathbb{R}^4}[\int_{\mathbb{R}\times\mathbb{R}^{*}}\overline{\langle\hat{\psi}_{\gamma,E_0}|\hat{U}^{\gamma,E_0}(\theta,b,a,v)\hat{\phi}_{\gamma,E_0}\rangle}\rho(\gamma,E_0)dE_{0}d\gamma}\\
&&\times\int_{\mathbb{R}\times\mathbb{R}^{*}}\langle\hat{\psi}_{\gamma^{\prime},E_{0}^{\prime}}|\hat{U}^{\gamma^{\prime},E_{0}^{\prime}}(\theta,b,a,v)\hat{\phi}_{\gamma^{\prime},E_{0}^{\prime}}\rangle\rho(\gamma^{\prime},E_{0}^{\prime})dE_{0}^{\prime}d\gamma^{\prime}]\;d\theta\; db\; da\; dv\\
=\lefteqn{\frac{1}{N^2}\int_{(E_{0},\gamma)}\int_{(E_{0}^{\prime},\gamma^{\prime})}[\int_{\mathbb{R}^{4}}\{\int_{k\in\mathbb{R}}\int_{k^{\prime}\in\mathbb{R}}e^{-i(\gamma-\gamma^{\prime})\theta}e^{i(\gamma k-\gamma^{\prime}k^{\prime})a}e^{\frac{-i}{2m}(\gamma k^{2}-\gamma^{\prime}k^{\prime 2})b}}\\
&&\lefteqn{\times e^{-i(E_{0}-E_{0}^{\prime})b} \overline{\hat{\phi}_{\gamma,E_{0}}(k-mv)}\hat{\psi}_{\gamma,E_{0}}(k)\overline{\hat{\psi}_{\gamma^{\prime},E_{0}^{\prime}}(k^{\prime})}\hat{\phi}_{\gamma^{\prime},E_{0}^{\prime}}(k^{\prime}-mv)\;dk\;dk^{\prime}\}}\\
&&\times d\theta\;db\;da\;dv]\rho(\gamma,E_{0})\rho(\gamma^{\prime},E_{0}^{\prime})(dE_{0}\;d\gamma)(dE_{0}^{\prime}\;d\gamma^{\prime})\\
\\
=\lefteqn{\frac{2\pi}{N^2}\int_{(E_{0},\gamma)}\int_{(E_{0}^{\prime},\gamma^{\prime})}\delta(\gamma-\gamma^{\prime})[\int_{\mathbb{R}^{3}}\{\int_{k}\int_{k^{\prime}}e^{i(\gamma k-\gamma^{\prime}k^{\prime})a}e^{-\frac{i}{2m}(\gamma k^{2}-\gamma^{\prime}k^{\prime 2})b}e^{-i(E_{0}-E_{0}^{\prime})b}}\\
&&\times\overline{\hat{\phi}_{\gamma,E_{0}}(k-mv)}\hat{\psi}_{\gamma,E_{0}}(k)\overline{\hat{\psi}_{\gamma^{\prime},E_{0}^{\prime}}(k^{\prime})}\hat{\phi}_{\gamma^{\prime},E_{0}^{\prime}}(k^{\prime}-mv)\;dk\;dk^{\prime}\}db\;da\;dv]\\
&&\times\rho(\gamma,E_{0})\rho(\gamma^{\prime},E_{0}^{\prime})(dE_{0}\;d\gamma)(dE_{0}^{\prime}\;d\gamma^{\prime})\\
=\lefteqn{\frac{2\pi}{N^2}\int_{(E_{0},\gamma)}\int_{E_{0}^{\prime}\in\mathbb{R}}[\int_{\mathbb{R}^{3}}\{\int_{k}\int_{k^{\prime}}e^{i\gamma( k-k^{\prime})a}e^{-\frac{i\gamma}{2m}( k^{2}-k^{\prime 2})b}e^{-i(E_{0}-E_{0}^{\prime})b}}\\
&&\times\overline{\hat{\phi}_{\gamma,E_{0}}(k-mv)}\hat{\psi}_{\gamma,E_{0}}(k)\overline{\hat{\psi}_{\gamma,E_{0}^{\prime}}(k^{\prime})}\hat{\phi}_{\gamma,E_{0}^{\prime}}(k^{\prime}-mv)\;dk\;dk^{\prime}\}db\;da\;dv]\\
&&\times\rho(\gamma,E_{0})\rho(\gamma,E_{0}^{\prime})dE_{0}\ d\gamma\ dE_{0}^{\prime}\\
=\lefteqn{\frac{(2\pi)^2}{N^2}\int_{(E_{0},\gamma)}\int_{E_{0}^{\prime}\in\mathbb{R}}[\int_{\mathbb{R}^{2}}\{\int_{k}\int_{k^{\prime}}\frac{\delta(k-k^{\prime})}{|\gamma|}e^{-\frac{i\gamma}{2m}( k^{2}-k^{\prime 2})b}e^{-i(E_{0}-E_{0}^{\prime})b}}\\
&&\times\overline{\hat{\phi}_{\gamma,E_{0}}(k-mv)}\hat{\psi}_{\gamma,E_{0}}(k)\overline{\hat{\psi}_{\gamma,E_{0}^{\prime}}(k^{\prime})}\hat{\phi}_{\gamma,E_{0}^{\prime}}(k^{\prime}-mv)\;dk\;dk^{\prime}\}db\;dv]\\
&&\times\rho(\gamma,E_{0})\rho(\gamma,E_{0}^{\prime})dE_{0}d\gamma dE_{0}^{\prime}\\
=\lefteqn{\frac{(2\pi)^2}{N^2}\int_{(E_{0},\gamma)}\int_{E_{0}^{\prime}\in\mathbb{R}}[\int_{\mathbb{R}^{2}}\{\int_{k}e^{-i(E_{0}-E_{0}^{\prime})b}\overline{\hat{\phi}_{\gamma,E_{0}}(k-mv)}\hat{\psi}_{\gamma,E_{0}}(k)\overline{\hat{\psi}_{\gamma,E_{0}^{\prime}}(k)}}\\
&&\times\hat{\phi}_{\gamma,E_{0}^{\prime}}(k-mv)\;dk\}db\;dv]\frac{\rho(\gamma,E_{0})}{|\gamma|}\rho(\gamma,E_{0}^{\prime})dE_{0}\ d\gamma\ dE_{0}^{\prime}\\
=\lefteqn{\frac{(2\pi)^3}{N^2}\int_{(E_{0},\gamma)}\int_{E_{0}^{\prime}\in\mathbb{R}}[\int_{v\in\mathbb{R}}\{\int_{k}\delta(E_{0}-E_{0}^{\prime})\overline{\hat{\phi}_{\gamma,E_{0}}(k-mv)}\hat{\psi}_{\gamma,E_{0}}(k)\overline{\hat{\psi}_{\gamma,E_{0}^{\prime}}(k)}}\\
&&\times\hat{\phi}_{\gamma,E_{0}^{\prime}}(k-mv)\;dk\}dv]\frac{\rho(\gamma,E_{0})}{|\gamma|}\rho(\gamma,E_{0}^{\prime})dE_{0} d\gamma dE_{0}^{\prime}\\
=\lefteqn{\frac{(2\pi)^3}{N^2}\int_{(E_{0},\gamma)}[\int_{v\in\mathbb{R}}\{\int_{k}\overline{\hat{\phi}_{\gamma,E_{0}}(k-mv)}\hat{\psi}_{\gamma,E_{0}}(k)\overline{\hat{\psi}_{\gamma,E_{0}}(k)}\hat{\phi}_{\gamma,E_{0}}(k-mv)\;dk\}}\\
&&\times dv]\frac{[\rho(\gamma,E_{0})]^2}{|\gamma|} d\gamma dE_{0}\\
=\lefteqn{\frac{(2\pi)^3}{N^2}\int_{(E_{0},\gamma)}\frac{[\rho(\gamma,E_{0})]^2}{|\gamma|} d\gamma dE_{0}[\int_{v\in\mathbb{R}}\{\int_{k}\overline{\hat{\phi}_{\gamma,E_{0}}(k-mv)}\hat{\psi}_{\gamma,E_{0}}(k)\overline{\hat{\psi}_{\gamma,E_{0}}(k)}}\\
&&\times\hat{\phi}_{\gamma,E_{0}}(k-mv)\;dk\}dv]\\
=\lefteqn{\frac{(2\pi)^3}{N^2}\int_{(E_{0},\gamma)}\frac{[\rho(\gamma,E_{0})]^2}{|\gamma|} d\gamma dE_{0}[\frac{1}{m}\int_{k^{\prime}\in\mathbb{R}}\{\int_{t\in\mathbb{R}}\overline{\hat{\phi}_{\gamma,E_{0}}(t)}\hat{\psi}_{\gamma,E_{0}}(k^{\prime})\overline{\hat{\psi}_{\gamma,E_{0}}(k^{\prime})}}\\
&&\times\hat{\phi}_{\gamma,E_{0}}(t)\;dt\}dk^{\prime}]\\
=\lefteqn{\frac{(2\pi)^3}{N^2}\int_{(E_{0},\gamma)}\frac{[\rho(\gamma,E_{0})]^2}{m|\gamma|}\|\hat{\phi}\|^{2}\|\hat{\psi}\|^{2} d\gamma dE_{0}.}\\
\end{eqnarray*}
Now, in view of (\ref{eq:ortho cond}) and (\ref{eq:ortho-con-presnt-secnario}), we finally obtain (in momentum space)
\begin{eqnarray}\label{eq:duflo-moore-planch-quant-gal}
\rho(\gamma,E_0)&=&m|\gamma|\\
N&=& (2\pi)^{3/2},
\end{eqnarray}
which are the Plancherel measure and the Duflo-Moore operator, respectively, for the quantum Galilei group case.

\subsection{Computation that leads to the fact that the canonical exponent $\xi_{1}$ used to construct $\mathcal{G}^{m}$ is no good to compute the correct Wigner function}\label{subsec:can-exp-wigfunc}
The most general expression for Wigner function is given by the following expression \cite{plancherel}
\begin{eqnarray}\label{eq:most-genrl-expr-wigfnc-plnch}
\lefteqn{W(A|X_{\lambda}^*)=\frac{[\sigma_{\lambda}(X_{\lambda}^*)]^{\frac{1}{2}}}{(2\pi)^{\frac{n}{2}}}\int_{N_{0}}e^{-i\langle  X_{\lambda}^{*};X\rangle}}\nonumber\\
&&\times[\int_{\hat{G}}\mbox{tr}(U_{\sigma}(e^{-X})[A(\sigma)C_{\sigma}^{-1}])[m(X)]^{\frac{1}{2}}d\nu_{G}(\sigma)]dX.
\end{eqnarray}
The first exponential term in (\ref{eq:most-genrl-expr-wigfnc-plnch}) is given by the following expression
\begin{equation*}
\exp i(k_{1}^{*}v-k_{2}^{*}a+\{k_1+\frac{(k_{2}^{*})^2}{2mk_2}\}b+k_{2}\theta).
\end{equation*}
And the densities $\sigma$ and $m$ for the (1+1) dimensional Galilei group turn out to be simply 1. Here, we are interested in the coadjoint orbits $\mathcal{O}^{*}_{k_1,k_2}$ due to dual algebra elements $\begin{bmatrix}0&0&k_1&k_2\end{bmatrix}$.
The induced representation of $\mathcal{G}^{m}$ was found to be
\begin{equation*}
(\hat{U}^{\gamma,E_0}(\theta,b,a,v)\hat{\phi})(k)=e^ {i[\gamma(\theta+\frac{k^{2}b}{2m}-ka)+E_{0}b]}\hat{\phi}(k-mv).
\end{equation*}
It follows immediately that
\begin{eqnarray}\label{eq:rep-modfd-quant-gal}
\lefteqn{(\hat{U}^{\gamma,E_{0}}(e^{-Y})\hat{\phi})(k)=(\hat{U}^{\gamma,E_{0}}(-\theta,-b,a,-v)^{-1}\hat{\phi})(k)}\nonumber\\
&&=(\hat{U}^{\gamma,E_{0}}(\theta+\frac{1}{2}mv^{2}b-mva,b,vb-a,v)\hat{\phi})(k)\nonumber\\
&&=e^{i\gamma\theta+\frac{i}{2}\gamma mv^{2}b-i\gamma mva+i\gamma\frac{k^{2}b}{2m}-i\gamma kvb+i\gamma ka+iE_{0}b}\hat{\phi}(k-mv).
\end{eqnarray}
The Duflo-Moore operator was found to be just $(2\pi)^{3/2}$. So $C^{-1}$ in (\ref{eq:most-genrl-expr-wigfnc-plnch}) is just $\frac{1}{(2\pi)^{3/2}}$. Also, in this case $n=4$. Now combining (\ref{eq:rep-modfd-quant-gal}) with the exponential term mentioned at the start and putting them in (\ref{eq:most-genrl-expr-wigfnc-plnch}), we find the following
\begin{eqnarray}\label{eq:wigfnc-quant-gal-nogood}
\lefteqn{W(|\phi_{\gamma,E_0}\rangle\langle\psi_{\gamma,E_0}|;k_{1}^{*},k_{2}^{*};k_1,k_2)}\nonumber\\
&&=\frac{m}{(2\pi)^{3}\sqrt{2\pi}}\int_{\theta}\int_{b}\int_{a}\int_{v}e^{ i(k_{1}^{*}v-k_{2}^{*}a+\{k_1+\frac{(k_{2}^{*})^2}{2mk_2}\}b+k_{2}\theta)}\nonumber\\
&&\times[\int_{\gamma}\int_{E_{0}}\int_{k}e^{i\gamma\theta-i\gamma kvb+i\gamma ka+\frac{i}{2}\gamma mv^{2}b-i\gamma mva+i\gamma\frac{k^{2}b}{2m}+iE_{0}b}
\overline{\hat{\psi}_{\gamma,E_{0}}(k)}\hat{\phi}_{\gamma,E_{0}}(k-mv)\nonumber\\
&&\times|\gamma|dk\;dE_{0}\;d\gamma]dv\;da\;db\;d\theta\nonumber\\
&&=\frac{m|k_2|}{(2\pi)^{2}\sqrt{2\pi}}\int_{E_{0}}[\int_{b}\int_{a}\int_{v}\int_{k}e^{ ik_{1}^{*}v-ik_{2}^{*}a}e^{i\{k_1+\frac{(k_{2}^{*})^2}{2mk_2}\}b}\nonumber\\
&&\times\overline{\hat{\psi}_{-k_{2},E_{0}}(k)}e^{ik_{2}kvb-ik_{2}ka-\frac{i}{2}k_{2}mv^{2}b+ik_{2}mva-ik_{2}\frac{k^{2}b}{2m}+iE_{0}b}\hat{\phi}_{-k_{2},E_{0}}(k-mv)\nonumber\\
&&\times dk\;dv\;da\;db]dE_{0}\nonumber\\
&&=\frac{m|k_2|}{(2\pi)^2\sqrt{2\pi}}\int_{E_0}\int_{b}\int_{k}\int_{v}[\int_{a}e^{-ik_{2}^{*}a-ik_{2}ka+ik_{2}mva}da]e^{ik_{1}^{*}v}e^{i\{k_{1}+\frac{(k_{2}^{*})^{2}}{2mk_{2}}\}b}\nonumber\\
&&\times\overline{\hat{\psi}_{-k_{2},E_{0}}(k)}e^{ik_{2}kvb-\frac{i}{2}k_{2}mv^{2}b-ik_{2}\frac{k^{2}b}{2m}+iE_{0}b}\hat{\phi}_{-k_{2},E_{0}}(k-mv)\;dv\;dk\;db\;dE_{0}\nonumber\\
&&=\frac{m|k_2|}{2\pi\sqrt{2\pi}}\int_{E_0}\int_{b}\int_{k}\int_{v}\delta(k_{2}k+k_{2}^{*}-k_{2}mv)e^{ik_{1}^{*}v}e^{i\{k_{1}+\frac{(k_{2}^{*})^{2}}{2mk_{2}}\}b}\nonumber\\
&&\times\overline{\hat{\psi}_{-k_{2},E_{0}}(k)}e^{ik_{2}kvb-\frac{i}{2}k_{2}mv^{2}b-ik_{2}\frac{k^{2}b}{2m}+iE_{0}b}\hat{\phi}_{-k_{2},E_{0}}(k-mv)\;dv\;dk\;db\;dE_{0}\nonumber\\
&
&=\frac{m}{2\pi\sqrt{2\pi}}\int_{E_0}\int_{b}\int_{v}e^{ik_{1}^{*}v}e^{ik_{1}b+iE_{0}b}\overline{\hat{\psi}_{-k_{2},E_{0}}(-\frac{k_{2}^{*}}{k_{2}}+mv)}\hat{\phi}_{-k_{2},E_{0}}(-\frac{k_{2}^{*}}{k_{2}})\nonumber\\
&&\times db\;dv\;dE_{0}\nonumber\\
&&=\frac{m}{\sqrt{2\pi}}\int_{v}e^{ik_{1}^{*}v}\overline{\hat{\psi}_{-k_{2},-k_1}(-\frac{k_{2}^{*}}{k_{2}}+mv)}\hat{\phi}_{-k_{2},-k_1}(-\frac{k_{2}^{*}}{k_{2}})dv.
\end{eqnarray}
Thus, we do not arrive at the correct Wigner function using the canonical exponent $\xi_{1}$.
\subsection{Coadjoint action matrix and UIRs of the (1+1)-centrally extended Galilei group $\mathcal{G}^{m\prime}$}\label{subsec:coad-act-mat-uir-new-exp}
The geometry of the coadjoint orbits for the quantum Galilei group $\mathcal{G}^{m}$ is encoded in the coadjoint action matrix given by (\ref{eq:coad-act-mat-quant-galgrp}). Now, if we extend the (1+1)-Galilei group $\mathcal{G}_{0}$ using the exponent $\xi_{2}$ (see (\ref{eq:new-expt})) to yield $\mathcal{G}^{m\prime}$, the geometry of the coadjoint orbits of $\mathcal{G}^{m\prime}$ remains unaltered, when compared to those of $\mathcal{G}^{m}$. Now to verify that we will compute the coadjoint action matrix for $\mathcal{G}^{m\prime}$ explicitly.

A generic lie algebra element is denoted as $Y=a_{1}K+a_{2}X+a_{3}T+a_{4}\Theta$ where $K,X,T,\Theta$ are the lie algebra generators corresponding to boost, space translation, time translation and the central extension respectively. They are given by the following matrices
\begin{equation*}
K=\begin{bmatrix}0&0&0&0\\0&0&0&-1\\\frac{1}{2}m&0&0&0\\0&0&0&0\end{bmatrix},\;
X=\begin{bmatrix}0&0&0&1\\0&0&0&0\\0&\frac{1}{2}m&0&0\\0&0&0&0\end{bmatrix},\;
T=\begin{bmatrix}0&1&0&0\\0&0&0&0\\0&0&0&0\\0&0&0&0\end{bmatrix},\;
\Theta=\begin{bmatrix}0&0&0&0\\0&0&0&0\\0&0&0&1\\0&0&0&0\end{bmatrix}.
\end{equation*}
And the corresponding commutation relations are exactly the one that we obtained using the canonical exponent $\xi_{1}$
\begin{eqnarray*}
&&[K,X]= m\Theta,\;[K,T]= X,\;[X,T]= 0\\
&&[\Theta,T]= 0,\;[\Theta,K]= 0,\;[\Theta,X]= 0.
\end{eqnarray*}
So, in terms of the above generators a generic lie algebra element reads
\begin{equation}\label{eq:lie-algbr-elmt-new-expt}
Y=\begin{bmatrix}0&a_{3}&0&a_{2}\\0&0&0&-a_{1}\\\frac{1}{2}ma_{1}&\frac{1}{2}ma_{2}&0&a_{4}\\0&0&0&0\end{bmatrix}
\end{equation}
But a generic group element $(\theta,b,a,v)$ is given by the following matrix
\begin{equation}\label{eq:lie-grp-elmt-new-expt}
(\theta,b,a,v)=\begin{bmatrix}1&b&0&a-vb\\0&1&0&-v\\\frac{1}{2}mv&\frac{1}{2}ma
&1&\theta\\0&0&0&1\end{bmatrix},
\end{equation}
obeying the group multiplication rule given by (\ref{eq:grp-law-ex-gal-new}).

Also, $(\theta,b,a,v)^{-1}=(-\theta,-b,vb-a,-v)$. The matrix representation of an inverse group element follows
\begin{equation}\label{eq:invrs-grp-elmt-new-expt}
(\theta,b,a,v)^{-1}=\begin{bmatrix}1&-b&0&-a\\0&1&0&v\\\frac{1}{2}mv&\frac{1}{2}m(vb-a)&1&-\theta\\0&0&0&1\end{bmatrix}
\end{equation}
Now, given the fact that a generic group element is given by equation (\ref{eq:lie-grp-elmt-new-expt}) and its inverse by (\ref{eq:invrs-grp-elmt-new-expt}) and that the adjoint action of a group element on the lie algebra element is defined by $\hbox{Ad}_{g}Y=gYg^{-1}$, we have

\begin{eqnarray*}
\lefteqn{\hbox{Ad}_{g^{-1}}(Y)}\\
&&=\begin{bmatrix}0&a_{3}&0&ba_{1}+a_{2}-a_{3}v\\0&0&0&-a_1\\\frac{1}{2}ma_{1}&\frac{1}{2}m(ba_{1}+a_{2}-va_{3})&0&ma_{1}(a-vb)-mva_{2}+\frac{1}{2}ma_{3}v^{2}+a_{4}\\0&0&0&0\end{bmatrix}.
\end{eqnarray*}
From which the coadjoint action matrix for $\mathcal{G}^{m\prime}$ follows as
 \begin{equation}\label{eq:coad-act-mat-new-expt}
M(g^{-1})=\begin{bmatrix}1&0&0&0\\b&1&-v&0\\0&0&1&0\\m(a-vb)&-mv&\frac{1}{2}mv^{2}&1\end{bmatrix}.
\end{equation}
The above matrix is exactly the same as found in (\ref{eq:coad-act-mat-quant-galgrp}).
\subsection{Computation of the Wigner function for $\mathcal{G}^{m\prime}$}\label{eq:wigfnc-new-expt}
The unitary irreducible representations of the (1+1)-centrally extended Galilei group $\mathcal{G}^{m}$ or the quantum Galilei group was computed in (\ref{eq:uir-quant-gal}). The central extension procedure was carried out by the canonical exponent introduced in (\ref{eq:can-expt}). The other exponent $\xi_2$ yielding the centrally extended group $\mathcal{G}^{m\prime}$ is mentioned in (\ref{eq:new-expt}). The two exponents introduced are equivalent in the sense of \cite{bargmann}. In other words, the difference between the two exponents is a trivial one, which can be written by means of the following continuous function
\begin{equation}\label{eq:con-fnc-def-triv-expt}
\zeta_{T}{(b,a,v)}=\frac{mva}{2},
\end{equation}
in the following way
\begin{eqnarray*}
\lefteqn{\xi_{1}(b_1,a_1,v_1)-\xi_{2}(b_2,a_2,v_2)}\\
&&=\frac{1}{2}mv_{1}a_{2}+\frac{1}{2}mv_{2}a_{1}+\frac{1}{2}mv_{1}^{2}b_{2}+\frac{1}{2}mv_{1}v_{2}b_{2}\\
&&=\zeta_{T}((b_1,a_1,v_1)(b_2,a_2,v_2))-\zeta_{T}(b_1,a_1,v_1)-\zeta_{T}(b_2,a_2,v_2),
\end{eqnarray*}
which entails the fact that $\hat{U}^{\prime\;\gamma,E_{0}}:=e^{\frac{i\gamma mva}{2}}\hat{U}^{\gamma,E_{0}}$ would be a projective representation of the (1+1)-Galilei group $\mathcal{G}_{0}$, where $\gamma$ is introduced for dimensional consistency and for keeping track with (\ref{eq:uir-quant-gal}). In the language of ordinary representations, we can state that $\hat{U}^{\prime\;\gamma,E_{0}}$, so obtained, is a unitary irreducible representation of the (1+1)-centrally extended Galilei group $\mathcal{G}^{m\prime}$. The fact that irreducibility is preserved during the whole process of arriving at a unitary representation (e.g., $\hat{U}^{\prime\;\gamma,E_{0}}$) of the central extension of the given group (e.g., $\mathcal{G}_{0}$) with resepect to a certain multiplier (e.g., using the canonical exponent $\xi_{1}$) from the known UIR (e.g., $\hat{U}^{\gamma,E_{0}}$) of a central extension of the same group corresponding to another multiplier (e.g., the one due to $\xi_{2}$) by means of projecting and lifting it in several steps is described in \cite{Projective}.

Therefore, the UIRs of the (1+1)-centrally extended Galilei group $\mathcal{G}^{m\prime}$ acting on $L^{2}(\hat{\mathbb{R}},dk)$ is given by
\begin{equation}\label{eq:uirs-new-expt}
(\hat{U}^{\prime\;\gamma,E_{0}}(\theta,b,a,v)\hat{\phi})(k)=\exp i[\gamma(\theta-ka+\frac{mva}{2}+\frac{k^{2}b}{2m})+E_{0}b]\hat{\phi}(k-mv).
\end{equation}

Now, following the steps as mentioned in section \ref{subsec:duflo-moore-planch-meas-quant-gal}, we can obtain the Duflo-Moore operator and the Plancherel measure for $\mathcal{G}^{m\prime}$ which are as follows
\begin{eqnarray}\label{eq:duflo-moore-plancrl-new-expt}
\rho(\gamma,E_0)&=&m|\gamma|\\
N&=& (2\pi)^{3/2}.
\end{eqnarray}
These are exactly the same as obtained for the other central extension of the (1+1)-Galilei group.

Now, we are well-equipped to compute the Wigner function for the centrally extended group $\mathcal{G}^{m\prime}$.
 If $Y$ ia a generic Lie algebra element given by $y=a_{1}K-a_{2}X+a_{3}T+a_{4}\Theta$, it follows from (\ref{eq:uirs-new-expt}) (under the identification $a_{1}\leftrightarrow -v, a_{2}\leftrightarrow -a, a_3\leftrightarrow -b, a_{4}\leftrightarrow -\theta$) that
\begin{eqnarray*}
(\hat{U}^{\gamma,E_{0}}(e^{Y})\hat{\phi})(k)&=&(\hat{U}^{\gamma,E_{0}}(-\theta,-b,a,-v)\hat{\phi})(k)\\
&=&e^{i[\gamma\{-\theta-ka-\frac{mva}{2}-\frac{k^{2}b}{2m}\}-E_{0}b]}\hat{\phi}(k+mv)
\end{eqnarray*}
Therefore,
\begin{eqnarray*}
(\hat{U}^{\gamma,E_{0}}(e^{-Y})\hat{\phi})(k)&=&(\hat{U}^{\gamma,E_{0}}(-\theta,-b,a,-v)^{-1}\hat{\phi})(k)\\
&=&(\hat{U}^{\gamma,E_{0}}(\theta,b,vb-a,v)\hat{\phi})(k)\\
&=&e^{i\gamma\theta-i\gamma kvb+i\gamma ka+\frac{i}{2}\gamma mv^{2}b-\frac{i}{2}\gamma mva+i\gamma\frac{k^{2}b}{2m}+iE_{0}b}\hat{\phi}(k-mv).
\end{eqnarray*}

We are going to use the most general expression for the Wigner function given in (\ref{eq:most-genrl-expr-wigfnc-plnch}) to compute the Wigner function of $\mathcal{G}^{m\prime}$ like we used to compute that of $\mathcal{G}^{m}$.

Here also, both the densities $\sigma$ and $m$ are simply 1. So, the Wigner function for the the (1+1)-centrally extended Galilei group $\mathcal{G}^{m\prime}$ reads

\begin{subequations}
\begin{align}
\MoveEqLeft W(|\phi_{\gamma,E_0}\rangle\langle\psi_{\gamma,E_0}|;k_{1}^{*},k_{2}^{*};k_1,k_2)\nonumber\\
&=\frac{m}{(2\pi)^{3}\sqrt{2\pi}}\int_{\theta}\int_{b}\int_{a}\int_{v}e^{ i(k_{1}^{*}v-k_{2}^{*}a+\{k_1+\frac{(k_{2}^{*})^2}{2mk_2}\}b+k_{2}\theta)}\nonumber\\
&\times[\int_{\gamma}\int_{E_{0}}\int_{k}e^{i\gamma\theta-i\gamma kvb+i\gamma ka+\frac{i}{2}\gamma mv^{2}b-\frac{i}{2}\gamma mva+i\gamma\frac{k^{2}b}{2m}+iE_{0}b}
\overline{\hat{\psi}_{\gamma,E_{0}}(k)}\hat{\phi}_{\gamma,E_{0}}(k-mv)\nonumber\\
&\times|\gamma|dk\;dE_{0}\;d\gamma]dvdadbd\theta\nonumber\\
&=\frac{m|k_2|}{(2\pi)^{2}\sqrt{2\pi}}\int_{E_{0}}[\int_{b}\int_{a}\int_{v}\int_{k}e^{ ik_{1}^{*}v-ik_{2}^{*}a}e^{i\{k_1+\frac{(k_{2}^{*})^2}{2mk_2}\}b}\overline{\hat{\psi}_{-k_{2},E_{0}}(k)}\nonumber\\
&\times e^{ik_{2}kvb-ik_{2}ka-\frac{i}{2}k_{2}mv^{2}b+\frac{ik_{2}mva}{2}-ik_{2}\frac{k^{2}b}{2m}+iE_{0}b}\hat{\phi}_{-k_{2},E_{0}}(k-mv)\;dk\;dv\;da\;db]dE_{0}\nonumber\\
&=\frac{m|k_2|}{(2\pi)^{2}\sqrt{2\pi}}\int_{E_0}\int_{b}\int_{k}\int_{v}[\int_{a}e^{-ik_{2}^{*}a-ik_{2}ka+\frac{ik_{2}mva}{2}}da]e^{ik_{1}^{*}v}e^{i\{k_{1}+\frac{(k_{2}^{*})^{2}}{2mk_{2}}\}b}\overline{\hat{\psi}_{-k_{2},E_{0}}(k)}\nonumber\\
&\times e^{ik_{2}kvb-\frac{i}{2}k_{2}mv^{2}b-ik_{2}\frac{k^{2}b}{2m}+iE_{0}b}\hat{\phi}_{-k_{2},E_{0}}(k-mv)\;dv\;dk\;db\;dE_{0}\nonumber\\
&=\frac{m|k_2|}{2\pi\sqrt{2\pi}}\int_{E_0}\int_{b}\int_{k}\int_{v}\delta(k_{2}k+k_{2}^{*}-\frac{k_{2}mv}{2})e^{ik_{1}^{*}v}e^{i\{k_{1}+\frac{(k_{2}^{*})^{2}}{2mk_{2}}\}b}\nonumber\\
&\times\overline{\hat{\psi}_{-k_{2},E_{0}}(k)}e^{ik_{2}kvb-\frac{i}{2}k_{2}mv^{2}b-ik_{2}\frac{k^{2}b}{2m}+iE_{0}b}\hat{\phi}_{-k_{2},E_{0}}(k-mv)\;dv\;dk\;db\;dE_{0}\nonumber\\
&=\frac{m}{2\pi\sqrt{2\pi}}\int_{E_0}\int_{b}\int_{v}e^{ik_{1}^{*}v}e^{i\{k_{1}+\frac{(k_{2}^{*})^{2}}{2mk_{2}}\}b}\overline{\hat{\psi}_{-k_{2},E_{0}}(-\frac{k_{2}^{*}}{k_{2}}+\frac{mv}{2})}\nonumber\\
&\times\hat{\phi}_{-k_{2},E_{0}}(-\frac{k_{2}^{*}}{k_{2}}+\frac{mv}{2}-mv)e^{ik_{2}vb(-\frac{k_{2}^{*}}{k_2}+\frac{mv}{2})-\frac{i}{2}k_{2}mv^{2}b-\frac{ik_{2}b}{2m}(-\frac{k_{2}^{*}}{k_{2}}+\frac{mv}{2})^{2}+iE_{0}b}\nonumber\\
&\times dv\;db\;dE_{0}\nonumber\\
&=\frac{m}{2\pi\sqrt{2\pi}}\int_{E_0}\int_{b}\int_{v}e^{ik_{1}^{*}v}e^{ik_{1}b-\frac{i}{2}k_{2}^{*}vb-\frac{i}{8}k_{2}mv^{2}b+iE_{0}b}\overline{\hat{\psi}_{-k_{2},E_{0}}(-\frac{k_{2}^{*}}{k_{2}}+\frac{mv}{2})}\nonumber\\
&\times\hat{\phi}_{-k_{2},E_{0}}(-\frac{k_{2}^{*}}{k_{2}}-\frac{mv}{2})\;db\;dv\;dE_{0}\nonumber\\
&=\frac{m}{2\pi\sqrt{2\pi}}\int_{E_0}\int_{v}e^{ik_{1}^{*}v}[\int_{b}e^{-i(-k_{1}+\frac{1}{2}k_{2}^{*}v+\frac{1}{8}k_{2}mv^{2}-E_{0})b}db]\overline{\hat{\psi}_{-k_{2},E_{0}}(-\frac{k_{2}^{*}}{k_{2}}+\frac{mv}{2})}\nonumber\\
&\times\hat{\phi}_{-k_{2},E_{0}}(-\frac{k_{2}^{*}}{k_{2}}-\frac{mv}{2})dvdE_{0}\label{eq:wignc-lead-weyl-heisen}\\
&=\frac{m}{\sqrt{2\pi}}\int_{E_0}\int_{v}e^{ik_{1}^{*}v}\delta(E_{0}+k_{1}-\frac{k_{2}^*v}{2}-\frac{k_{2}mv^{2}}{8})\overline{\hat{\psi}_{-k_{2},E_{0}}(-\frac{k_{2}^{*}}{k_{2}}+\frac{mv}{2})}\nonumber\\
&\times\hat{\phi}_{-k_{2},E_{0}}(-\frac{k_{2}^{*}}{k_{2}}-\frac{mv}{2})dv\;dE_{0}\nonumber\\
&=\frac{m}{\sqrt{2\pi}}\int_{v}e^{ik_{1}^{*}v}\overline{\hat{\psi}_{-k_{2},-k_{1}+\frac{k_{2}^{*}v}{2}+\frac{k_{2}mv^{2}}{8}}(-\frac{k_{2}^{*}}{k_{2}}+\frac{mv}{2})}\nonumber\\
&\times\hat{\phi}_{-k_{2},-k_{1}+\frac{k_{2}^{*}v}{2}+\frac{k_{2}mv^{2}}{8}}(-\frac{k_{2}^{*}}{k_{2}}-\frac{mv}{2})dv.\label{eq:wigfnc-ex-gal-new-expt}
\end{align}
\end{subequations}
Equation (\ref{eq:wigfnc-ex-gal-new-expt}) gives the Wigner function for $\mathcal{G}^{m\prime}$ due to a two dimensional coadjoint orbit embedded in $\mathbb{R}^{4}$, the dual algbera space. But the associated Wigner functions are no longer supported on the corresponding coadjoint orbits. Rather the support is spread out through a whole family of coadjoint orbits as will be discussed in some detail in the following section.
\section{Summary of the main results and  physical interpretation}\label{sec:summ-results}
In the first half of the paper, we studied the (1+1)-extended affine Gaillei group $\mathcal{G}^{m}_{\hbox{\tiny{aff}}}$, the quantum version of the (1+1)-affine Galilei group $\mathcal{G}_{\hbox{\tiny{aff}}}$. The four nondegenerate coadjoint orbits of $\mathcal{G}^{m}_{\hbox{\tiny{aff}}}$ were all found to be six dimensional spaces embedded in $\mathbb{R}^{6}$, the underlying dual algebra space. These were basically cotangent bundles over the four open free $3$-dimensional Mackey orbits described in Figure \ref{fig:figfirst}. Hence, the coadjoint orbits of $\mathcal{G}^{m}_{\hbox{\tiny{aff}}}$ are also open free in $\mathbb{R}^{6}$. Open free orbits have nice structural implications in representation theory. And the Wigner functions associated with each such coadjoint orbit were found employing the standard techniques developed in \cite{semidirect}. Finally, the domains of the four Wigner functions were studied. Two of the functions were found to be supported inside the relavant coadjoint orbits while the other two were found to have their support spread out through the  $\mathbb{R}^{6}$ half hyperplane to which they belong.

Now, we look for a possible physical interpretation of the results obtained for the (1+1)-extended affine Galilei group $\mathcal{G}^{m}_{\hbox{\tiny{aff}}}$. The two representation-space parameters in this context were $q$ and $E_{0}:=E-\frac{p^2}{2mq}$. Different signs of $q$ and $E_{0}$ determine which coadjoint orbit we are in or which UIR we are talking about, as outlined in Table \ref{tab:tablefirst}. The mass term $m$ stands for mass scale or mass unit \cite{mahara}, while $qm$ stands for the physical mass which changes under the action of the dilation parameters ($\sigma$ and $\tau$). In other words, each unitary irreducible representation of $\mathcal{G}^{m}_{\hbox{\tiny{aff}}}$ represents a nonrelativistic spinless particle of variable mass ($qm$) and internal energy ($E_{0}$). And the Wigner function of $\mathcal{G}^{m}_{\hbox{\tiny{aff}}}$ associated with a nonrelativistic spinless particle of variable positive mass and changing positive internal energy was found to be supported inside its coadjoint orbit. The Wigner function due to a nonrelativistic spinless particle of varible negative mass and changing positive internal energy was also found to be supported inside its coadjoint orbit. It turns out that requiring symmetry under the affine Galilei group $\mathcal{G}^{m}_{\hbox{\tiny{aff}}}$ leads to no physically interesting phenomenon. But mathematically the coadjoint orbits were nicely structured and two of the relevant Wigner functions were found to be supported inside the corresponding coadjoint orbits.

Since the requirement of symmetry under dilation parameters led to no physically interesting object, in the later half of the paper we demanded only Galilean invariance and hence worked with the quantum Galilei group $\mathcal{G}^{m}$ and its  variant $\mathcal{G}^{m\prime}$ (where the extension was executed using an equivalent multiplier).

The Wigner function for the (1+1)-extended Galilei group $\mathcal{G}^{m\prime}$ that we found in (\ref{eq:wigfnc-ex-gal-new-expt}), is basically a map between two direct integral Hilbert spaces given by
\begin{eqnarray*}
\lefteqn{W:\int_{(\gamma,E_0)\in\mathbb{R}^{*}\times\mathbb{R}}^{\oplus}\mathcal{B}_{2}(L^{2}(\hat{\mathcal{O}}_{\gamma,E_0},dk))m|\gamma|d\gamma dE_{0}}\\
&&\rightarrow\int_{(k_1,k_2)\in\mathbb{R}\times\mathbb{R}^{*}}^{\oplus}L^{2}(\mathcal{O}_{k_2,k_1}^{*},dk_{1}^{*}dk_{2}^{*})dk_{1}dk_{2}.
\end{eqnarray*}
For
\begin{eqnarray*}
\phi_{k_2,k_1},\psi_{k_2,k_1}&\in& L^2(\mathcal{O}_{k_{2},k_{1}}^{*},dk_{1}^{*}dk_{2}^{*})\\
\hbox{and}\; |\phi_{\gamma,E_0}\rangle\langle\psi_{\gamma,E_0}|&\in& \mathcal{B}_{2}^{\oplus}(L^{2}(\hat{\mathcal{O}}_{\gamma,E_0},dk)),
\end{eqnarray*}
where $\mathcal{B}_{2}$ denotes the space of Hilbert-Schmidt operators and $\mathcal{B}_{2}^{\oplus}$ represents the direct integral Hilbert space of measurable Hilbert-Schmidt operator fields.
The Wigner function found in (\ref{eq:wigfnc-ex-gal-new-expt}) was restricted to the coadjoint orbit $\mathcal{O}_{k_2,k_1}^{*}$. But the final expression for the Wigner function reveals the fact that it is no longer supported on that single coadjoint orbit. Rather it has its support concentrated on the collection of coadjoint orbits exhausting $\mathbb{R}^3$ hyperplanes perpendicular to the fourth axis W. For brevity, we ask the reader to go back to section \ref{subsec:coadorb-quant-gal}, where we explained the geometry of the relevant coadjoint orbits. In particular, we have a continuous collection of Wigner functions being supported in each such hyperplane characterized by the constant $W=k_2$. This hyperplane can be called a \textquotedblleft support plane\textquotedblright. In other words, to each hyperplane (corresponding to a fixed value of $k_2\in\mathbb{R}\setminus\{0\}$ and $k_1$ assuming all real values in $\mathbb{R}$), we attach all such Wigner functions each pertaining to a non-relativistic spinless particle due to a fixed value of $\gamma$ (the analog of $q$, i.e. the mass scale in the extended affine case) and a definite real internal energy, having the states to be square integrable functions on the coadjoint orbits exhausting the $\mathbb{R}^3$ hyperpalne in question. In the language of representations we say that, the Wigner function restricted to the coadjoint orbit $\mathcal{O}^{*}_{k_2,k_1}=T^{*}\hat{\mathcal{O}}_{k_2,k_1}$ gets its contribution from representations associated to all the parabolic orbits corresponding to the $\mathbb{R}^2$ plane given by $\gamma=\gamma^{\prime}=k_2$ in the $\gamma^{\prime}E^{\prime}p^{\prime}$ space. The corresponding coadjoint orbits exactly exhaust the $\mathbb{R}^3$-hyperplane given by $W=k_2$ embedded in $\mathbb{R}^4$. Therefore, each Wigner map is associated with representations $U^{\gamma,E_0}$ due to a fixed value of $\gamma$ but all possible real values of $E_0$.

It was also found in Section \ref{subsec:coadorb-quant-gal} that it is important to choose the appropriate multiplier to find the correct Wigner function for the relevant group. The appropriateness is defined by the fact that the multiplier should reduce to one given by the Weyl-Heisenberg group under proper substitution. In this sense, $\xi_{2}$ defined by (\ref{eq:new-expt}) was found to be appropriate in the context of the (1+1)-Galilei group.

Next, we ask the question whether  under suitable conditions the Wigner function for $\mathcal{G}^{m\prime}$ given by equation (\ref{eq:wigfnc-ex-gal-new-expt}) reduces to the standard one due to the Weyl-Heisenberg group. One thing that marks a distinction between the above two groups is that the Galilei group has a time translation parameter while the Weyl-Heisenberg group does not, the second distinguishing characteristic being that the irreducible unitary representations of the former  are parameterized by two constants namely $\gamma$ and $E_0$ while those of the latter  are parameterized only by a single constant $\gamma$. With the above two considerations, we insert two Delta like functions $\delta(b)$ and $\delta(E_0)$ following the substitution of $k_1$ to be zero in equation (\ref{eq:wignc-lead-weyl-heisen}) to derive the following
\begin{eqnarray}\label{eq:reduc-weyl-heisen}
\lefteqn{W(|\phi_{k_{2}}\rangle\langle\psi_{k_{2}}|;k_{1}^{*},k_{2}^{*};k_2)}\nonumber\\
&&=\frac{m}{2\pi\sqrt{2\pi}}\int_{E_0}\int_{v}e^{ik_{1}^{*}v}[\int_{b}e^{-i(\frac{k_{2}^{*}v}{2}-\frac{1}{8}k_{2}mv^{2}-k_{1}-E_{0})b}\delta(b)db]\nonumber\\
&&\times\overline{\hat{\psi}_{k_{2},E_{0}}(-\frac{k_{2}^{*}}{k_{2}}+\frac{mv}{2})}\hat{\phi}_{k_{2},E_{0}}(-\frac{k_{2}^{*}}{k_{2}}-\frac{mv}{2})\delta(E_0)dvdE_{0}\nonumber\\
&&=\frac{m}{2\pi\sqrt{2\pi}}\int_{E_0}\int_{v}e^{ik_{1}^{*}v}\delta(E_{0})\overline{\hat{\psi}_{k_{2},E_{0}}(-\frac{k_{2}^{*}}{k_{2}}+\frac{mv}{2})}\hat{\phi}_{k_{2},E_{0}}(-\frac{k_{2}^{*}}{k_{2}}-\frac{mv}{2})dv\;dE_{0}\nonumber\\
&&=\frac{m}{2\pi\sqrt{2\pi}}\int_{v}e^{ik_{1}^{*}v}\overline{\hat{\psi}_{k_{2},0}(-\frac{k_{2}^{*}}{k_{2}}+\frac{mv}{2})}\hat{\phi}_{k_{2},0}(-\frac{k_{2}^{*}}{k_{2}}-\frac{mv}{2})dv.
\end{eqnarray}
The corresponding coadjoint orbits are foliated along one of the two othogonal axes of $\mathbb{R}^4$, along which the coadjoint orbits of the (1+1)-centrally extended Galilei group were foliated (see Section \ref{subsec:coadorb-quant-gal}, where the geometry of its coadjoint orbits is discussed at length). The measurable fields of the Hilbert-Schmidt operators are now indexed only by $\gamma$. The total orbit space (union of the disjoint coadjoint orbits in question) is no longer a dense subspace of $\mathbb{R}^4$. It is just a set of measure zero (lower dimensional space) instead. But it fills out an $\mathbb{R}^3$ hyperplane embedded in $\mathbb{R}^4$. Each coadjoint orbit here is a cotangent bundle of a parabola characterized by the constant $k_2$, which is homeomorphic to $\mathbb{R}^2$, the coadjoint orbit for the Weyl-Heisenberg group. So, we find that the Wigner function for the (1+1) dimensional extended Galilei group, computed using an appropriate multiplier, reduces to the one for the Weyl-Heisenberg group under proper substitution.

The coadjoint action matrix for the (1+1)-centrally extended Galilei group turned out to be independent of the choice of multipliers (belonging to the same equivalence class) required to do the extension. In other words, we always end up with the same coadjoint action matrix no matter what multiplier we choose from the same equivalence class of the second cohomology group $H^{2}(\mathcal{G}_{0},\mathbb{U}(1))$.

It is interesting to observe that the coadjoint orbits of the (1+1)-extended affine Galilei group $\mathcal{G}^{m}_{\hbox{\tiny{aff}}}$ (see Figure \ref{fig:figfirst}) disintegrate into a continuous family of parabolas, each of which is parameterized by a specific value of the ordered pair $(\gamma,E_{0})$. These parabolas are just the dual orbits of the (1+1)-centrally extended Galilei group $\mathcal{G}^{m}$ or the quantum Galilei group. This disintegration of the dual orbits resolves the difficulty of a nonrelativistic spinless particle possessing variable mass, but as a price of the remedy, the beauty of the structure of the open free orbits gets lost. None of the Wigner functions associated with the massive nonrelativistic spinless particles under the symmetry of the Galilei group remains supported inside the corresponding phase space. In this sense, to earn physically meaningful object we had to sacrifice the associated mathematically elegant structure.

\end{document}